\documentclass{article}

\usepackage{PRIMEarxiv}

\usepackage[utf8]{inputenc} 
\usepackage[T1]{fontenc}    
\usepackage{hyperref}       
\usepackage{url}            
\usepackage{booktabs}       
\usepackage{amsfonts}       
\usepackage{nicefrac}       
\usepackage{microtype}      
\usepackage{lipsum}
\usepackage{graphicx}
\graphicspath{{media/}}     
\title{A Cross-Platform Benchmark for Interval Computation Libraries}

\author{
  Xuan Tang\\New York University
    \And
    Zachary Ferguson\\New York University
    \And
    Teseo Schneider\\University of Victoria
    \And
    Denis Zorin\\New York University
    \And
    Shoaib Kamil\\Adobe Research
    \And
    Daniele Panozzo\\New York University
}

\usepackage{amsmath}
\usepackage{xspace}
\usepackage{listings}
\usepackage{booktabs}
\usepackage{adjustbox}
\usepackage{color}
\usepackage{float}
\usepackage{soul}
\usepackage{amssymb}
\usepackage{pifont}
\usepackage{lipsum}
\usepackage{adjustbox}

\definecolor{dkgreen}{rgb}{0,0.6,0}
\definecolor{gray}{rgb}{0.5,0.5,0.5}
\definecolor{mauve}{rgb}{0.58,0,0.82}

\usepackage[dvipsnames]{xcolor}
\definecolor{txcolor}{RGB}{21, 85, 59}

\definecolor{zfcolor}{RGB}{255, 127, 0}

\newcommand{\xmark}{\ding{55}}%
\newcommand{\boost}{Boost\xspace}%
\newcommand{\nativeswitched}{\texttt{native\_switched}\xspace}
\newcommand{\predsucc}{\texttt{pred\_succ}\xspace}
\newcommand{\multiplicative}{\texttt{multiplicative}\xspace}
\newcommand{\rounddown}{ROUND DOWN\xspace}
\newcommand{\roundup}{ROUND UP\xspace}
\newcommand{\roundnear}{ROUND NEAR\xspace}

\begin{document}
\maketitle

\begin{abstract}

Interval computation is widely used to certify computations that use floating point operations to avoid pitfalls related to rounding error introduced by inaccurate operations. Despite its popularity and practical benefits, support for interval arithmetic is not standardized nor available in mainstream programming languages. 

We propose the first benchmark for interval computations, coupled with reference solutions computed with exact arithmetic, and compare popular C and C++ libraries over different architectures, operating systems, and compilers. The benchmark allows  identifying limitations in existing implementations, and provides a reliable guide on which library to use on each system. We believe that our benchmark will be useful for developers of future interval libraries, as a way to test the correctness and performance of their algorithms.

\end{abstract}


\section{Introduction}
\label{sec:introduction}

Interval computation allows performing floating-point operations with certifiable correctness, by accounting for rounding errors. Every floating-point number is replaced by a pair of numbers, representing an interval that contains the exact result of the computation, independently from the rounding. While this approach increases the cost and memory usage of computations, it is a staple for many algorithms in computer aided design, geometric computing, image processing, computer graphics, and scientific computing. For example, they are used for Boolean computation \cite{CGAL}, intersections between parametric patches \cite{Snyder1993Interval}, continuous collision detection \cite{Redon2002fast}, subdivision surfaces \cite{Zorin2000}, and precision manufacturing \cite{Tibken1999Quality}. More applications are discussed in the survey \cite{Kearfott1996Interval}.

While the formal correctness of interval computation has been proven \cite{snyder1992interval}, ensuring that an implementation of interval arithmetic is correct is a daunting task, as the proof relies on assumptions on the order of operations (which can be altered by the compiler or the reordering buffers on the CPU) and on a set of hardware assumptions on the ALUs, which are architecture-dependent. At the same time, users rely on interval computation to certify the correctness of their algorithm, assuming that the interval computation library is correct, which, as we will show in this paper, is not always true for specific combinations of compilers, operating systems, and architectures.

Because a formal proof for every hardware and software combination is impractical (requiring to adapt the proof at every new software or hardware update), we propose an experimental approach: we introduce a large benchmark of test expressions and real-world algorithms for which the exact answer is computed using exact computation. The benchmark can then be used to test existing implementations, and identify issues. We note that a library passing our benchmark problems might still contain errors, as our benchmarks do not exhaustively test all possible combinations of operations and operands.

We use our benchmark to evaluate four popular C/C++ interval libraries (filib, filib++, \boost, BIAS) for correctness, interval size, speed, and consistency. The  results are summarized in Table \ref{fig:teaser}.
The only library that is at the same time correct, consistent, portable, and has a reasonable speed is filib, which does not rely on using special hardware instructions to control the underlying rounding mode. 

We provide the complete source code and scripts to run our benchmark, and in addition we provide a CMake build system for using filib on Windows, Linux, and macOS operating systems with both x86 and ARM architectures. We believe that our benchmark will be a useful tool to continue to assess the correctness of existing interval libraries as new compilers and architectures are developed, and also to provide a standardized set of tests for developers of interval libraries. 

\begin{figure}
    \centering
    \begin{adjustbox}{width=0.975\textwidth}
\begin{tabular}{lrrrrrr}
\toprule
{} &       BOOST &   FILIB &NATIVE SWITCHED&   MULTIPLICATIVE&   PRED-SUCC & BIAS\\
\midrule
CORRECTNESS (ARITHMETIC) & \checkmark& \checkmark & \checkmark& \checkmark& \checkmark& \checkmark\\
CORRECTNESS (TRANSCENDENTAL) & \xmark& \checkmark & \checkmark& \checkmark& \checkmark& \checkmark\\
CORRECTNESS (COMPOSITE) & \xmark& \checkmark & \xmark& \checkmark& \checkmark& \xmark\\
INTERVAL SIZE (ARITHMETIC) & 1& 2& 1&3 &2 & 1\\
INTERVAL SIZE (TRANSCENDENTAL) &1& 2& 2& 2& 2 &1 \\
SPEED &6 & 3& 4& 1 & 2 & 5\\
CONSISTENCY & \xmark& \checkmark& \xmark& \checkmark&\xmark & \xmark \\
PORTABILITY & \checkmark& \checkmark& \xmark& \xmark& \xmark & \xmark\\
\bottomrule
\end{tabular}
\end{adjustbox}
    \caption{We introduce a benchmark for interval arithmetic computation and test it on four C/C++ libraries: filib, filib++ (including the \nativeswitched, \multiplicative, and \predsucc methods), \boost, and BIAS. We evaluate each library for their correctness, output interval size, speed, consistency, and portability. The table shows a summary of our benchmark where the numbers indicate a ranking from best (small) to worst (large).}
    \label{fig:teaser}
\end{figure}

\section{Background}
\label{sec:background}

In the past two decades, numerous interval arithmetic libraries have been developed in various languages. While the logic behind interval arithmetic has been explored in many works~\cite{principle2imp,BENHAMOU19971}, the actual implementations vary from library to library and may produce different results. 

\subsection{Hardware Rounding Mode Control}
Many modern programming languages comply with the IEEE 754 standard for implementing floating point datatypes, which supports different rounding rules (round to nearest, towards $\infty$, and towards $-\infty$)~\cite{ieeeFloating}. These rounding modes provide good lower and upper bounds on basic arithmetic operations.
Libraries like \boost~\cite{BOOST} or CGAL \cite{CGAL} use this functionality to build interval operations. Their implementation focuses on setting the correct rounding mode before calling the default math library~\cite{BOOSTINTERVAL}. Other libraries like BIAS/PROFIL~\cite{BIAS} and filib++~\cite{filib++} also use or include this implementation for basic algebraic operations. 

Such strategies work well for basic arithmetic operations but require a lot of care when computing transcendental functions, where many rounding changes need to happen to evaluate a single transcendental function~\cite{transcendentalIA64}. Some of them, like CGAL, sidestep the problem by not supporting transcendental operations.

\subsection{Software Implementations}
It is possible to avoid relying on hardware rounding mode support by using a pure software implementation. There are two main approaches.


\paragraph*{Multiplicative.}
While it is hard to obtain the exact floating point error of an expression, the relative error of single operations can be generalized~\cite{worstcase,formalVerification} since there are only a finite number of bits representing a number~\cite{weshoudknow}. 
Hence, one can carefully analyze the error to generate a number $\epsilon$ such that if the true result is $\alpha$ and the computed result is $\beta$,  $(1-\epsilon)\beta\leq\alpha\leq(1+\epsilon)\beta$ holds and $1-\epsilon$, $1+\epsilon$ can be exactly represented in floating-point.
Filib++, BIAS, and GAOL~\cite{GAOL} all provide such implementations, although the choice of $\epsilon$ varies depending on how the analysis is performed.

\paragraph*{Changing binary representation.}
Since nowadays all floating-point numbers implementations follow the IEEE 754 standard, one can deconstruct the binary representation of a number and directly change the result to obtain an interval~\cite{efficientandreliable}.
Filib and filib++ adopt this approach, by directly modifying the mantissa and exponent of a double, generating a reasonably-small interval without sacrificing performance.

\subsection{Other Implementations}

Some libraries rely on others as part of their implementation of interval arithmetic. For example, IBEX~\cite{IBEX} and XSC~\cite{XSC} both use filib as the back end for interval computation. These libraries generally do not provide better performance or smaller interval size, but they focus on providing a more user-friendly interface. 
Other interval libraries exist in other programming languages. For example, IntervalArithmetic.jl~\cite{JULIALIB} in Julia, interval-arithmetic~\cite{JAVASCRIPTLIB} in Javascript. Since our goal is on C/C++ libraries we do not include such libraries for our study.


\section{Methodology}
\label{sec:method}

A good interval library should maintain four traits: (1) correctness, (2) small interval sizes, (3) efficiency, and (4) consistency across different architectures and compilers.
We design our benchmark to test these four traits.
We recognize that in many applications, an interval itself is initialized from a single number rather than an actual range since the goal is to compute an interval that includes the true value of an expression. Hence, the initialization of an interval in our benchmark is always from a single value.
In our benchmark, we compare the following four popular open source libraries: filib, filib++, \boost, and BIAS. 

Filib++ supports three modes for interval computations: \nativeswitched (uses system rounding modes), \predsucc (directly manipulates the bit representation of a double), and \multiplicative (multiplies two numbers to generate an interval). 
BIAS includes three rounding modes (\rounddown, \roundup, and \roundnear) which can be set before an interval operation. Their documentation is unclear how an interval operation is affected by these rounding modes, thus we treat them as three different interval types.

\subsection{Correctness}
While libraries can optimize interval operations for every single arithmetic or transcendental function, composite expressions that combine multiple operations can potentially cause the library to produce incorrect (interval does not include the true result) or empty (lower bound is greater than upper bound) intervals.
In our benchmark, we test each library on 28 different basic expressions and 104 expressions from FPBench~\cite{fpbench}, a floating point accuracy benchmark that covers a variety of application domains. The basic benchmark is composed of: four basic arithmetic operations (addition, subtraction, multiplication, and division); four transcendental functions ($\operatorname{sqrt}$, $\exp$, $\sin$, $\cos$); ten composite expressions that only contain basic arithmetic operations; and ten composite expressions containing both arithmetic operations and transcendental functions. These expressions are randomly generated from a fixed seed and listed on the website.

For each expression, we generate one million valid inputs for evaluation. To ensure that the representation of the input and result are precise, and no additional floating point error is introduced during validation, we convert every input and output to rational format using GMP~\cite{GMP}.
A typical query has the form
\[
\frac{n_{l}}{d_{l}} \leq \operatorname{expression}(\frac{n_1}{d_1}, \ldots) \leq \frac{n_u}{d_u}
\]
for $n,d\in\mathbb{Z}$.
Using this format, the queries can be evaluated later by an arbitrary precision software to get an exact answer. In our benchmark, we use Mathematica \cite{Mathematica}.

\subsection{Interval size}
To report the interval size we utilize a similar procedure to when checking correctness. Instead of outputting the actual query, we compute the interval size by using a rational subtraction (i.e., we convert the upper and lower bound to rational numbers). 

\subsection{Speed}
To test the speed of an interval library, we measure the execution time for each expression: we generate 1{,}000 inputs for each expression and execute the expression 10{,}000 times for every input. Finally, we accumulate the total execution time for each library and expression to report the performance.
It is important to execute different sets of inputs since input values may affect the performance of some operations due to range reduction.

\subsection{Consistency and Portability}
We deployed our benchmark on four different platforms with different compilers: Windows (Intel Core i7 8700k, x86-64, Windows 10, MSVC 14.27.29110), macOS Intel (2.4 GHz 8-Core Intel Core i9, x86-64, macOS Big Sur, Darwin Kernel Version 20.1.0, Apple clang version 12.0.0), macOS Arm (3.2 GHz 4-Core/2 + 2 GHz 4-Core Apple M1, arm64, macOS Big Sur, Darwin Kernel Version 20.1.0, Apple clang version 12.0.0), Linux (AMD EPYC 7452 32-Core Processor, x86-64, Ubuntu 19.10, GCC 9.2.1). 
%

\section{Results}
\label{sec:result}

We discuss in detail how each interval type perform over different platform and expressions.

\subsection{Correctness}
As discussed before, we test each library on 28 (constructed by us) and 104 (extracted from FPBench) expressions. 
We check for correctness by ensuring that the interval computation produces an interval containing the exact solution, evaluated with arbitrary precision with Mathematica \cite{Mathematica}.

We begin with the 28 expressions. All of the libraries produce correct results for basic arithmetic operations. However, when it comes to transcendental functions, \boost is not correct (since it deals with transcendental functions by setting rounding modes before calling the standard math library). Specifically, it fails for $\exp$ and trigonometry functions, where the implementation is based on Taylor expansion~\cite{transcendentalIA64}. For composite expressions that only contain basic arithmetic operations, all libraries are correct. When transcendental functions are included in a composite expression, BIAS produces incorrect intervals for Expression~\eqref{expression:random1}. 

Filib and filib++'s three interval modes are correct for the 28 expressions, the \nativeswitched mode for filib++ is not correct on four of the expressions from FPBench. For example, ``polarToCarthesian, x'', that computes
\[
    r \cos(\theta\cdot(3.14159265359 / 180.0))
\]
which contains transcendental function $\cos$. Another example is the expression ``sineOrder3'' 
\[
    (0.954929658551372  x_0) - (0.12900613773279798  ((x_0 \, x_0) \, x_0))
\]
which only contains basic arithmetic operations, and is designed to find floating point problem caused by the order of evaluation.

We conclude that only filib and filib++'s \predsucc and \multiplicative modes produce correct intervals for all tests.

\subsection{Interval size}

\begin{figure*}
    \parbox{0.02\linewidth}{~}\hfill\hfill
    \parbox{0.24\linewidth}{\centering Windows}\hfill
    \parbox{0.24\linewidth}{\centering Mac}\hfill
    \parbox{0.24\linewidth}{\centering Linux}\hfill
    \parbox{0.24\linewidth}{\centering ARM}\par
    \parbox{0.02\linewidth}{\centering\rotatebox{90}{\sffamily\scriptsize{Percentage}}}\hfill\hfill
    \parbox{0.24\linewidth}{\centering\sffamily\scriptsize
    \includegraphics[width=\linewidth]{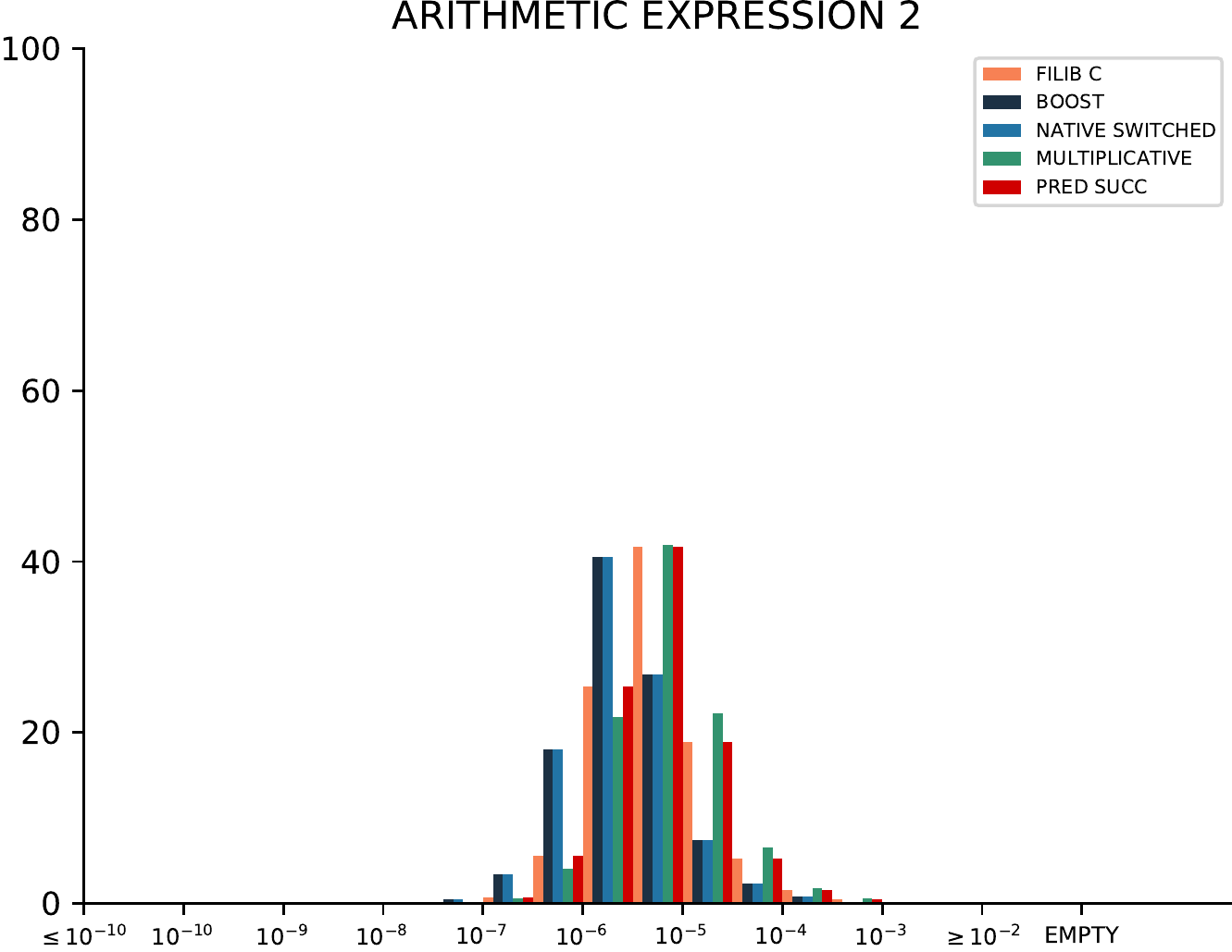}}\hfill
    \parbox{0.24\linewidth}{\centering\sffamily\scriptsize
    \includegraphics[width=\linewidth]{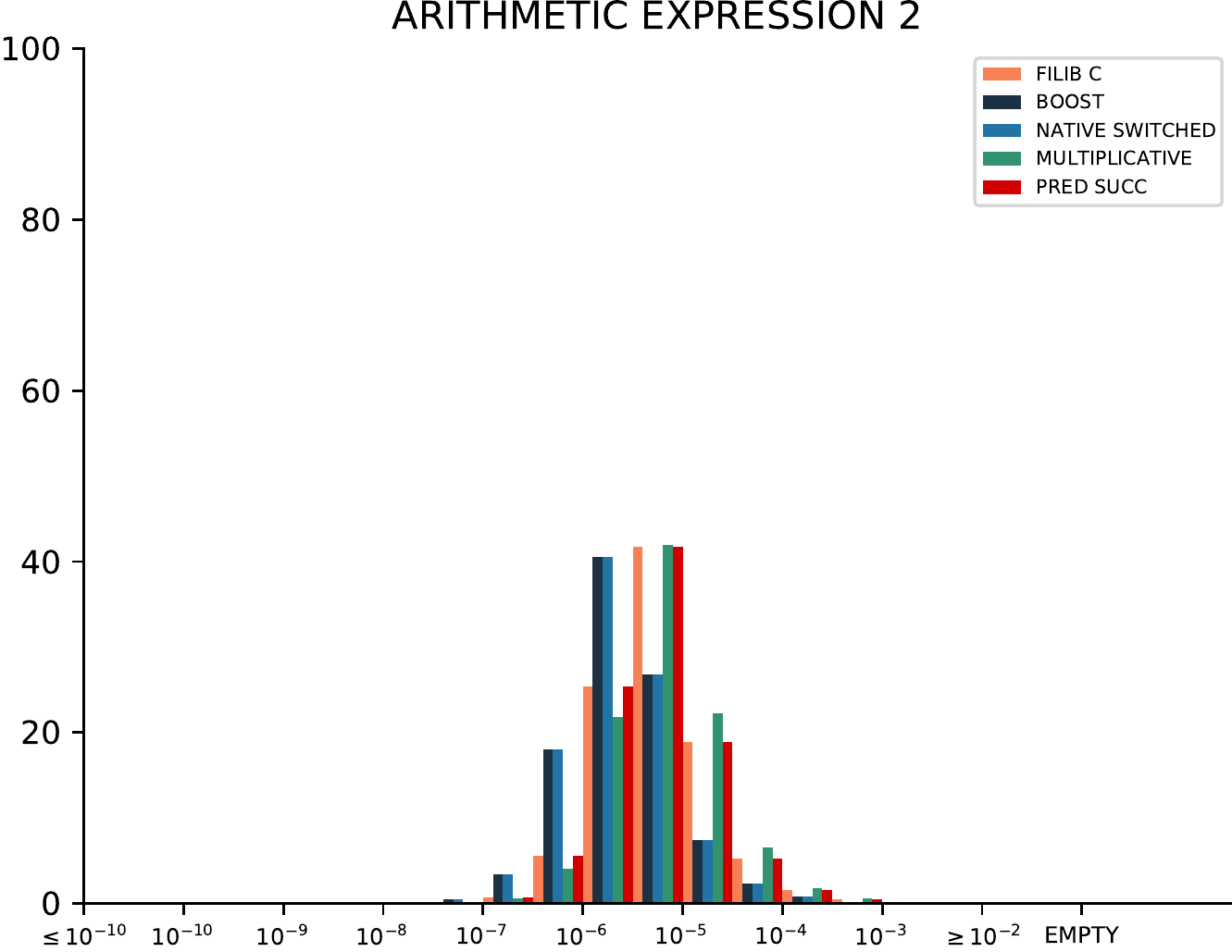}}\hfill
    \parbox{0.24\linewidth}{\centering\sffamily\scriptsize
    \includegraphics[width=\linewidth]{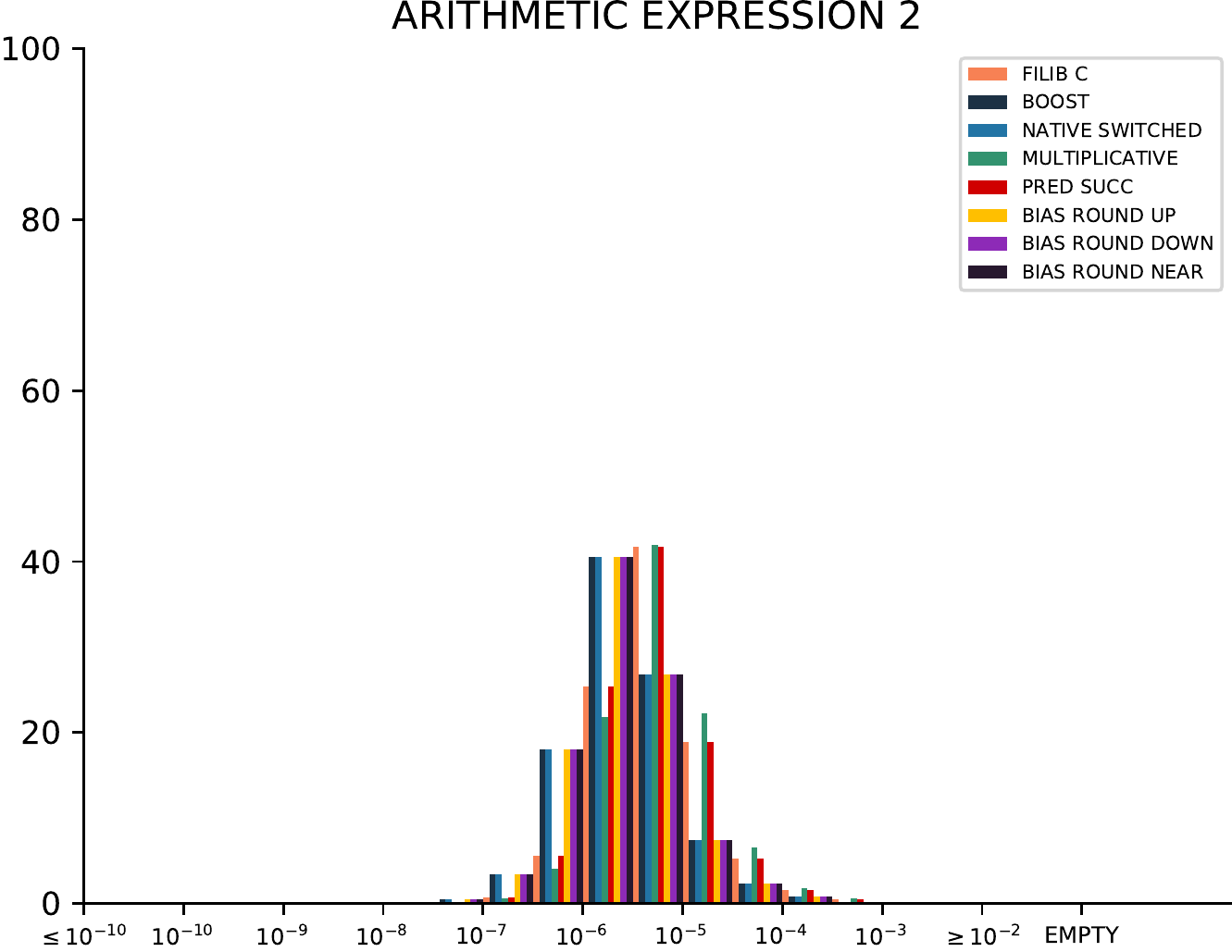}}\hfill
    \parbox{0.24\linewidth}{\centering\sffamily\scriptsize
    \includegraphics[width=\linewidth]{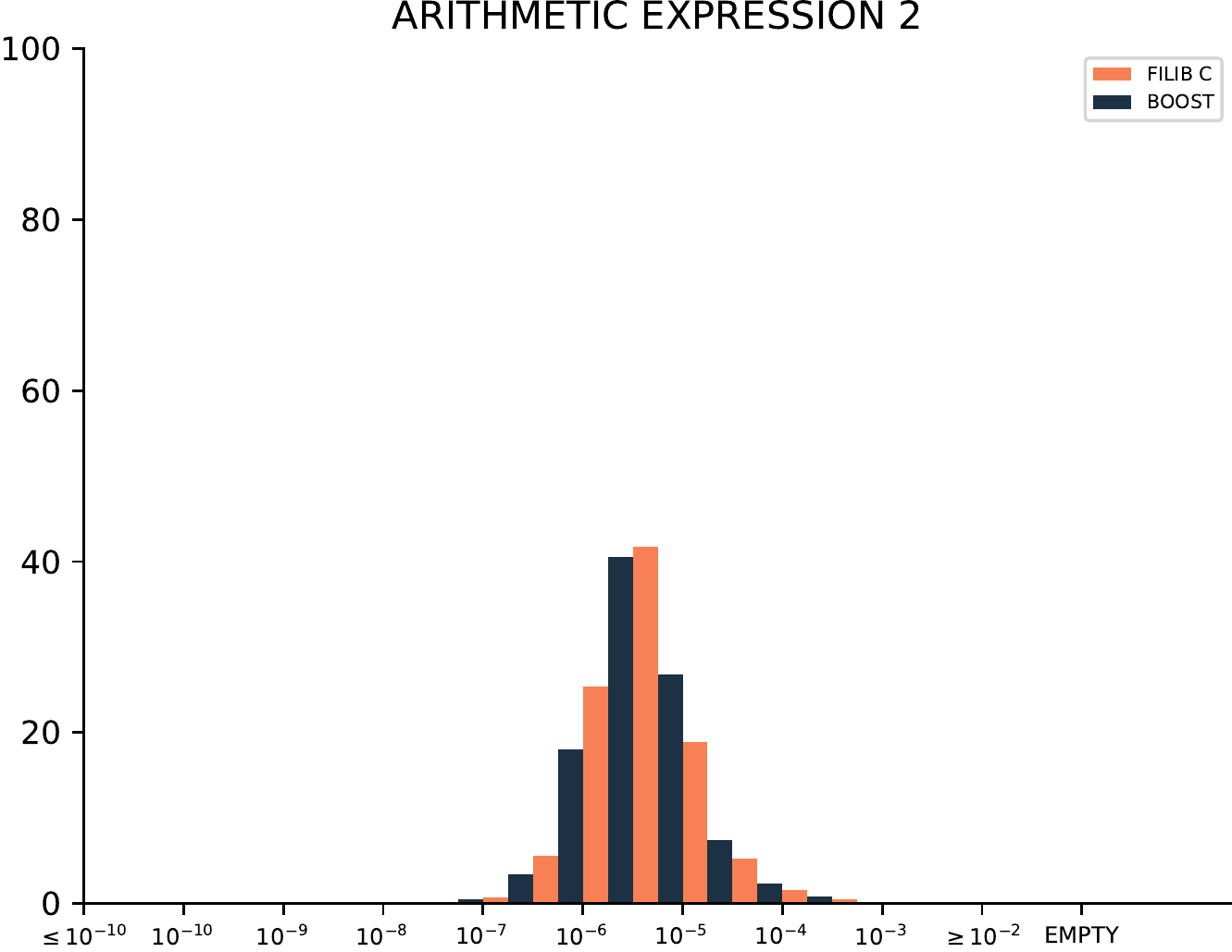}}\par
    \parbox{0.02\linewidth}{\centering\rotatebox{90}{\sffamily\scriptsize{Percentage}}}\hfill\hfill
    \parbox{0.24\linewidth}{\centering\sffamily\scriptsize
    \includegraphics[width=\linewidth]{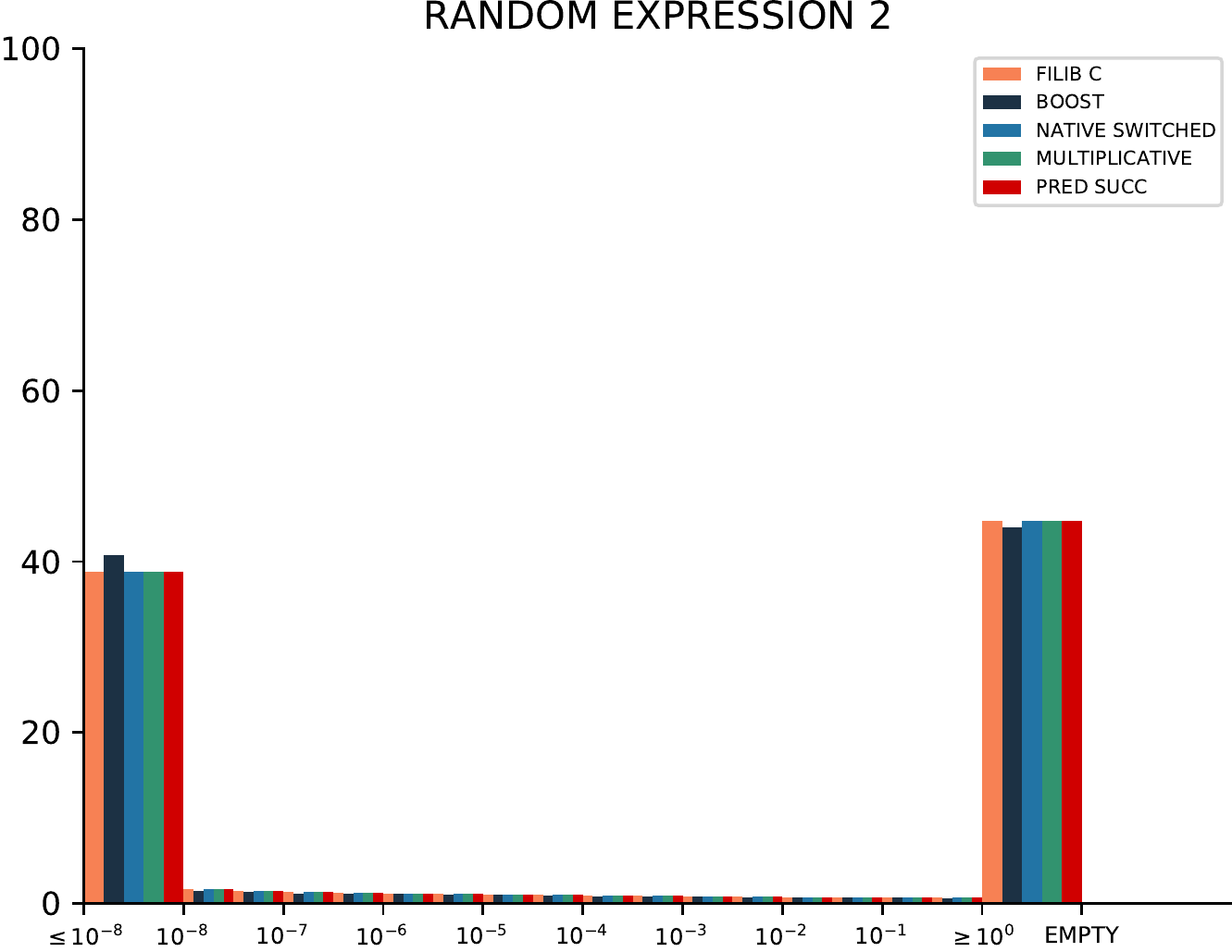}}\hfill
    \parbox{0.24\linewidth}{\centering\sffamily\scriptsize
    \includegraphics[width=\linewidth]{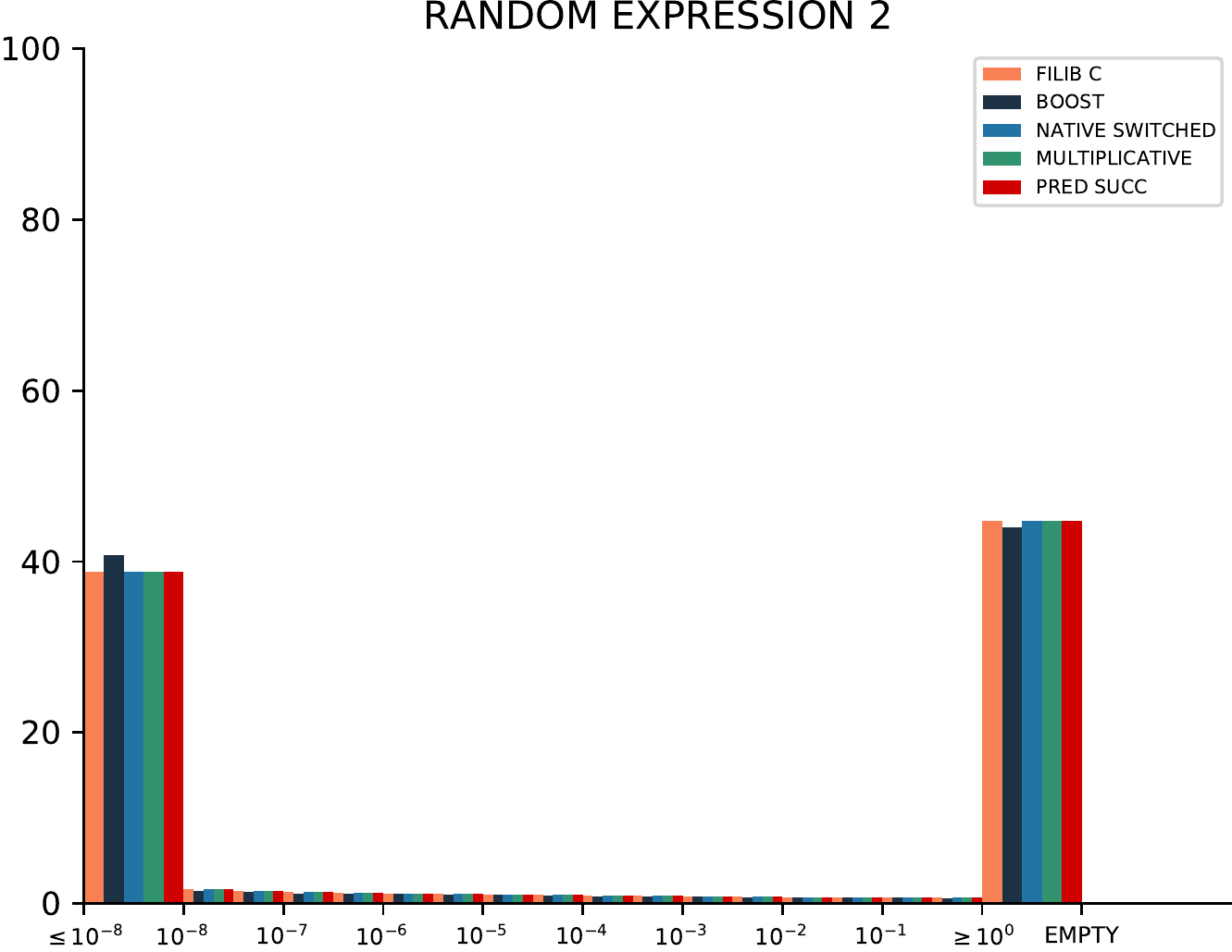}}\hfill
    \parbox{0.24\linewidth}{\centering\sffamily\scriptsize
    \includegraphics[width=\linewidth]{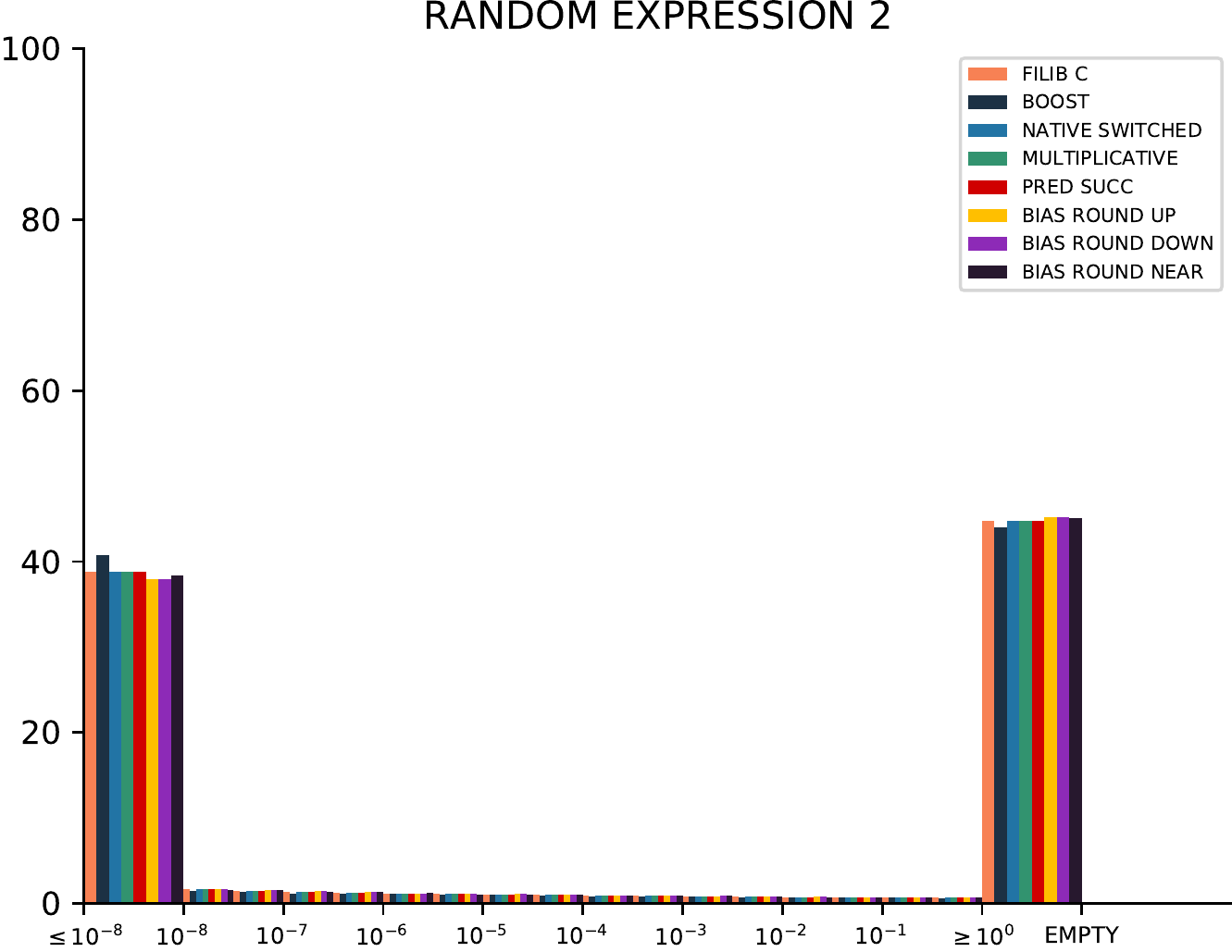}}\hfill
    \parbox{0.24\linewidth}{\centering\sffamily\scriptsize
    \includegraphics[width=\linewidth]{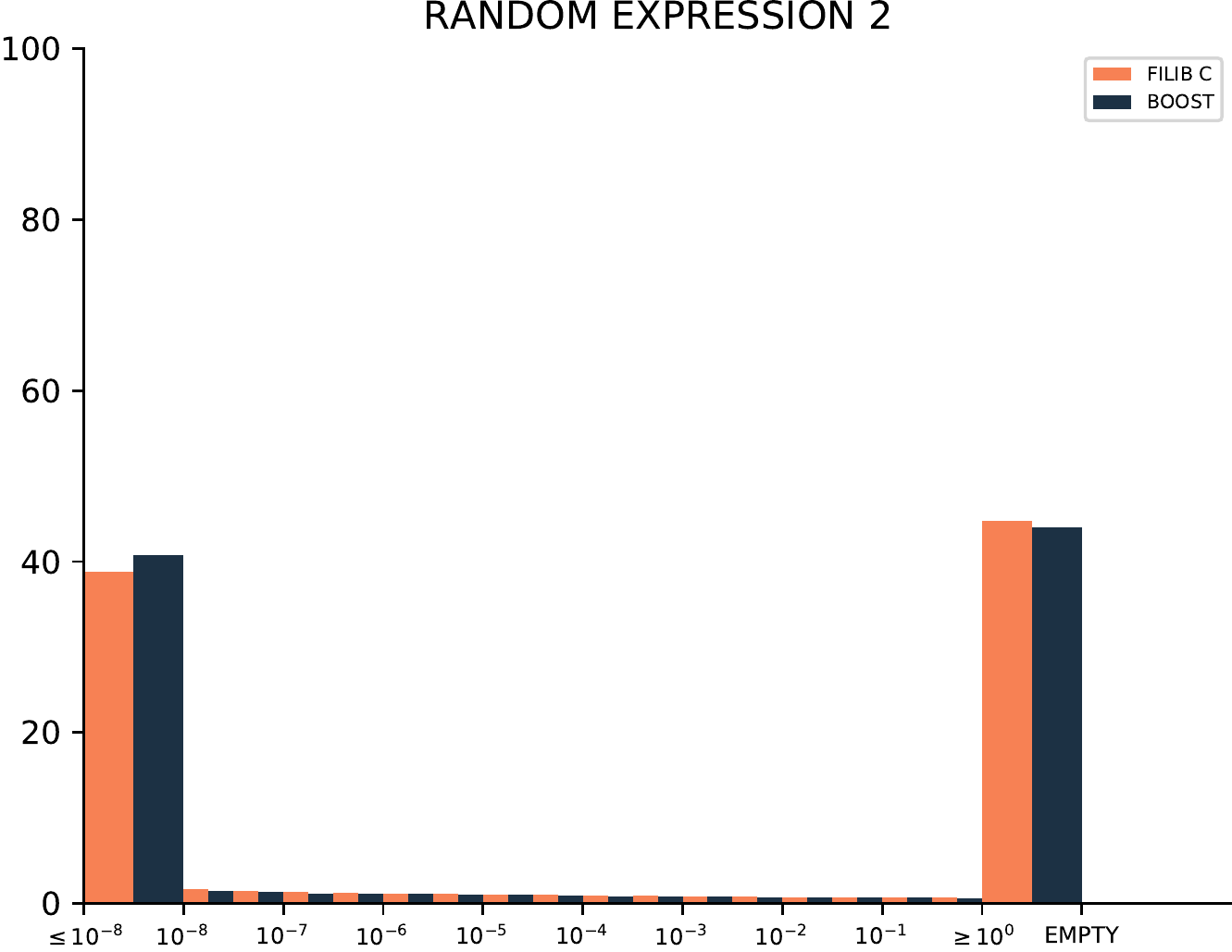}}\par
    \parbox{0.02\linewidth}{\centering\rotatebox{90}{\sffamily\scriptsize{Percentage}}}\hfill\hfill
    \parbox{0.24\linewidth}{\centering\sffamily\scriptsize
    \includegraphics[width=\linewidth]{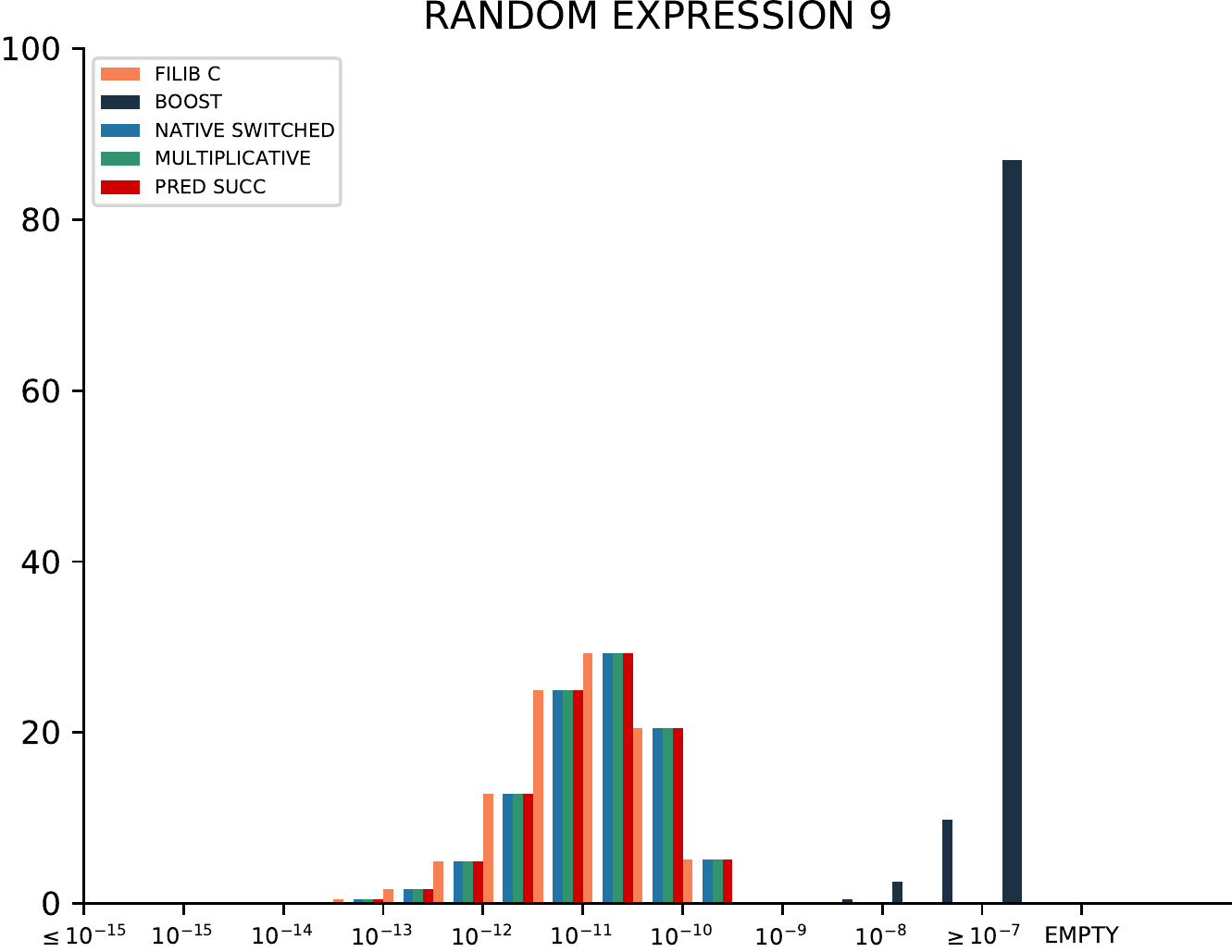}}\hfill
    \parbox{0.24\linewidth}{\centering\sffamily\scriptsize
    \includegraphics[width=\linewidth]{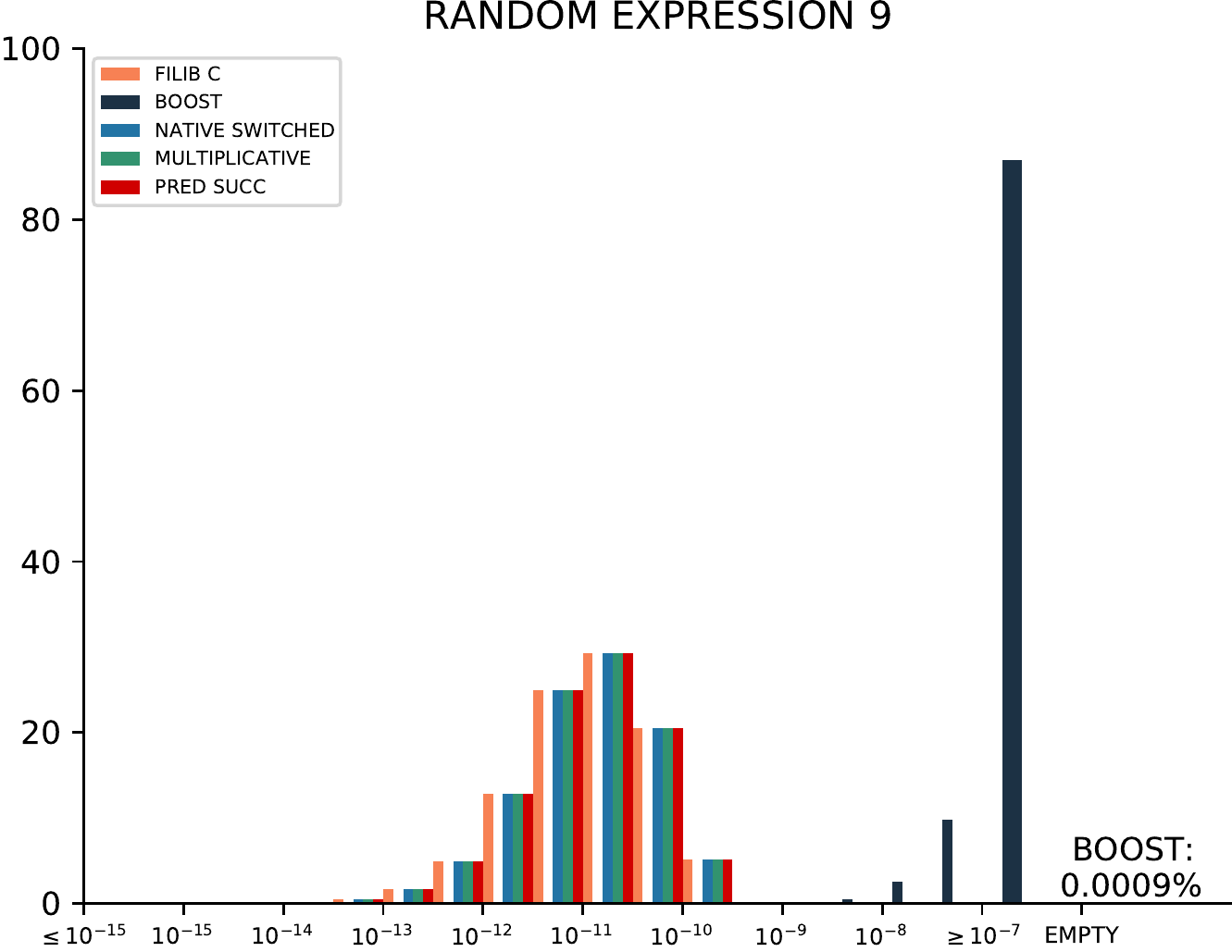}}\hfill
    \parbox{0.24\linewidth}{\centering\sffamily\scriptsize
    \includegraphics[width=\linewidth]{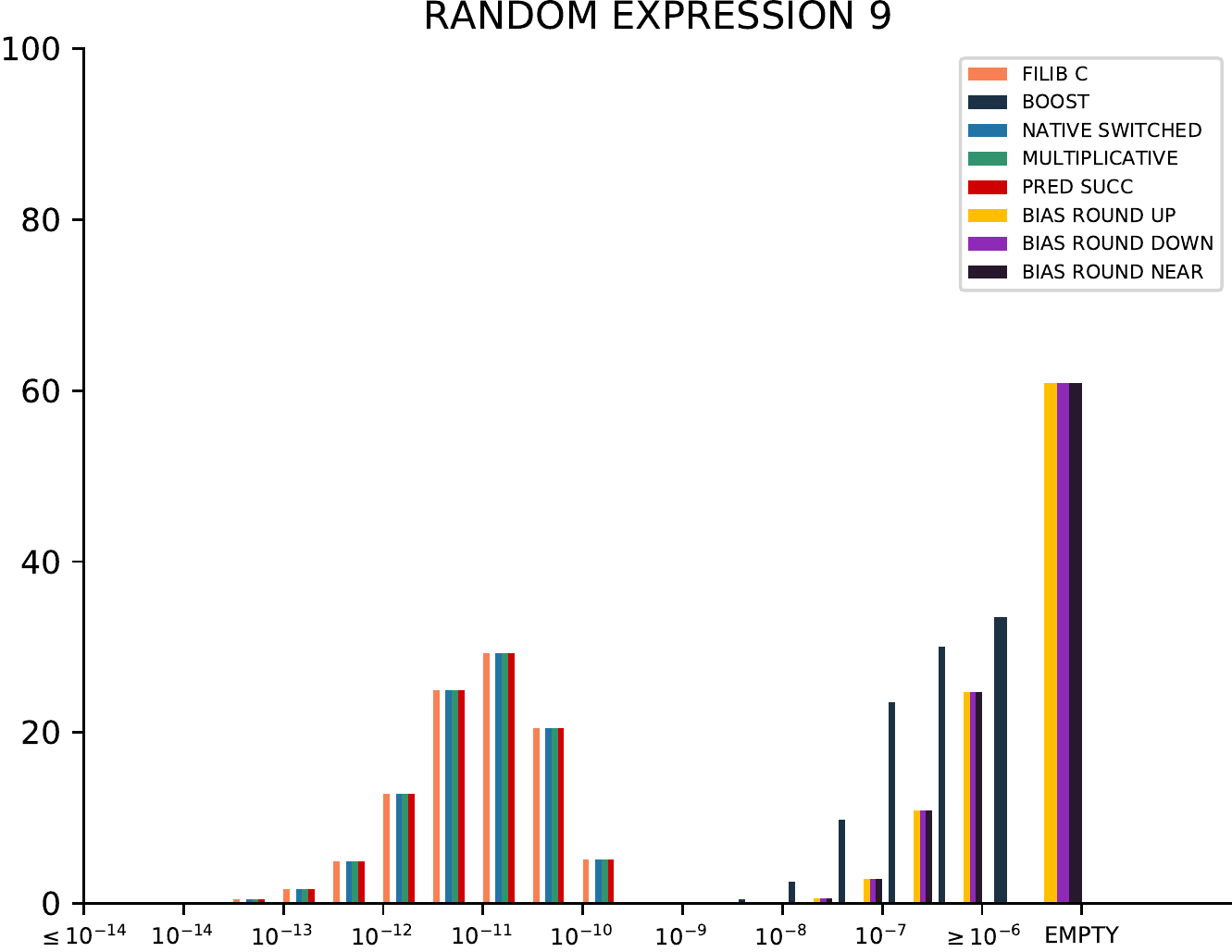}}\hfill
    \parbox{0.24\linewidth}{\centering\sffamily\scriptsize
    \includegraphics[width=\linewidth]{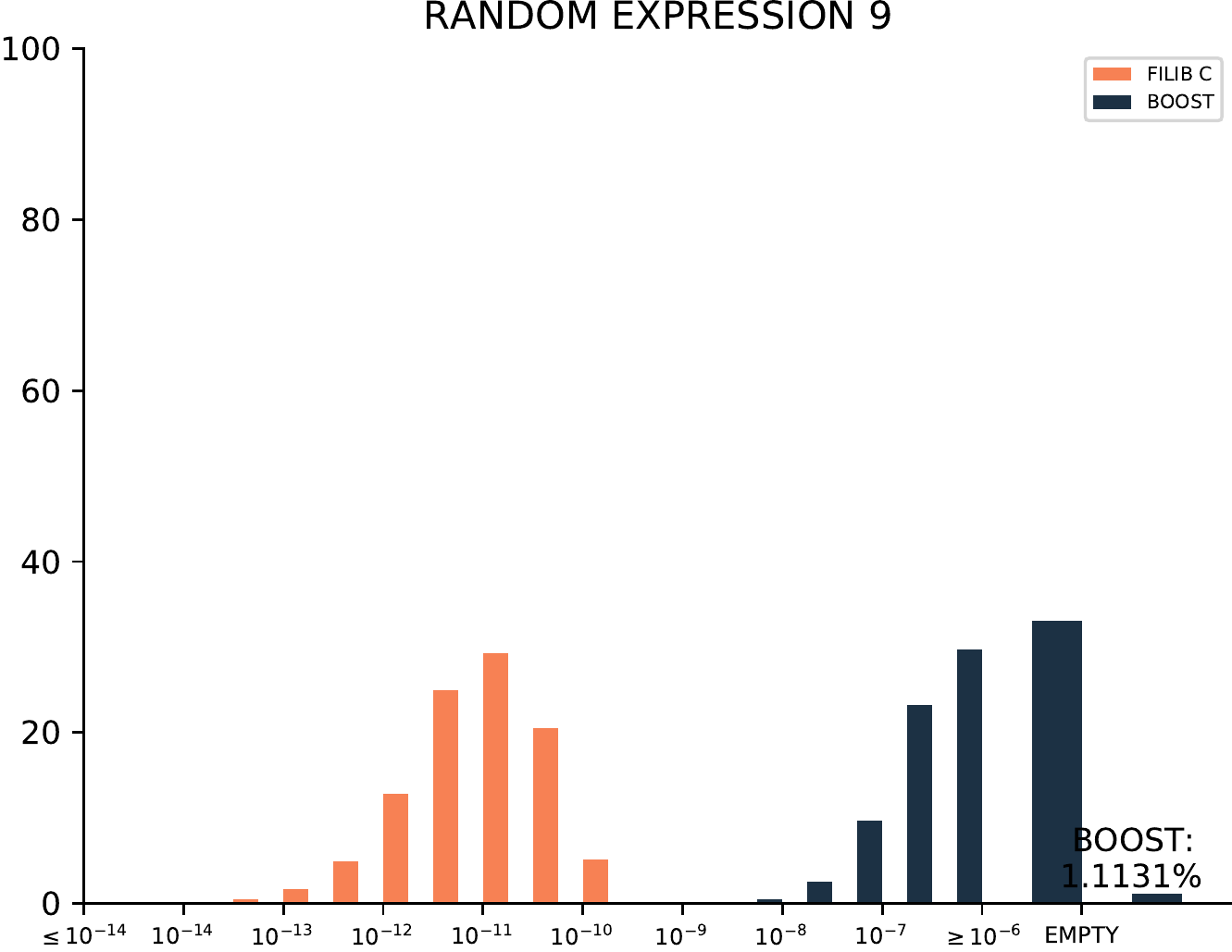}}\par
    \caption{Distribution of interval size. Top: Expression~\eqref{expression:arith2}. Middle: Expression~\eqref{expression:random2}. Bottom: Expression~\eqref{expression:random9}}
    \label{graph:gap_size}
\end{figure*}

Due to the large number of test expressions, we show only some of the most representative expressions. Specifically, we look at one expression that contains only arithmetic operations (Expression~\eqref{expression:arith2}), one expression that contains only transcendental functions (Expression~\eqref{expression:random2}), and one that contains both (Expression~\eqref{expression:random9}).

Across the different platforms, the distribution of interval size does not vary much. However, within each platform, the distribution of interval size can be quite different between libraries.
The top row of Figure~\ref{graph:gap_size} shows that for expressions that contain only arithmetic operations, libraries that use system rounding modes (\boost, filib++ \nativeswitched, BIAS) produce  smaller interval sizes compared to others. The \multiplicative mode of filib++ produces the largest interval sizes. However, the differences are small across libraries.

When transcendental functions are added into the expression, the interval sizes can be unpredictable (Figure \ref{graph:gap_size}). 
The difference of overall distribution of the libraries can also be quite large depending on the expression, but within filib and filib++'s three interval modes, the interval sizes are quite similar.
We also see that \boost produces empty intervals, an indication that \boost's results are sometimes incorrect.

\subsection{Performance}
\begin{table*}
    \centering
    \begin{adjustbox}{width=1\textwidth}
\begin{tabular}{lrrrrrrrr}
\toprule
{} &   FILIB C &     BOOST &   NATIVE SWITCHED &   MULTIPLICATIVE &   PRED SUCC &   BIAS UPWARD &   BIAS DOWNWARD &   BIAS NEAR \\
\midrule
 ADDITION                 &     66.52 &    405.78 &                    18.49 &                    3.94 &              {\textbf{3.92}} &        327.93 &          328.69 &      327.94 \\
 SUBTRACTION              &     80.08 &    405.35 &                    18.37 &                    9.20 &               {\textbf{3.90}} &        327.60 &          327.44 &      327.44 \\
 MULTIPLICATION           &     78.40 &    649.62 &                    25.81 &                   {\textbf{13.59}} &              90.03 &        339.43 &          338.87 &      338.61 \\
 DIVISION                 &     84.43 &    451.78 &                    {\textbf{38.13}} &                   68.97 &              81.12 &        344.25 &          343.87 &      344.13 \\
 SQUARE ROOT              &     84.85 &    434.21 &                    24.19 &                   24.26 &              {\textbf{24.18}} &         61.76 &           60.44 &       60.40 \\
 EXPONENTIAL              &    199.08 &    462.47 &                   {\textbf{143.28}} &                  143.43 &             143.43 &        196.44 &          196.34 &      200.39 \\
 SIN                      &    202.25 &   7115.77 &                   {\textbf{171.54}} &                  172.79 &             172.94 &       2088.65 &         2088.34 &     2087.43 \\
 COS                      &    190.21 &   6776.14 &                   168.64 &                  168.65 &             {\textbf{168.54}} &       2425.19 &         2424.21 &     2423.90 \\
 ARITEMETIC EXPRESSION 1  &    861.49 &   6848.71 &                  1969.52 &                  {\textbf{589.98}} &            1053.79 &       4126.49 &         4097.03 &     4091.17 \\
 ARITEMETIC EXPRESSION 2  &   1292.77 &  23227.45 &                  3170.62 &                  {\textbf{699.40}} &            1040.81 &       5706.45 &         5672.61 &     5661.78 \\
 ARITEMETIC EXPRESSION 3  &   1844.20 &  15082.05 &                  4984.85 &                 {\textbf{1318.50}} &            1900.36 &       7790.34 &         7769.46 &     7757.53 \\
 ARITEMETIC EXPRESSION 4  &   3758.37 &  30943.47 &                 11407.83 &                 {\textbf{2354.18}} &            4077.55 &      16086.02 &        16062.78 &    16054.29 \\
 ARITEMETIC EXPRESSION 5  &   2635.54 &  21860.23 &                  7910.95 &                 {\textbf{1626.65}} &            2888.93 &      10891.99 &        10883.80 &    10871.13 \\
 ARITEMETIC EXPRESSION 6  &   1254.13 &  20903.24 &                  3263.05 &                  {\textbf{651.20}} &            1022.87 &       5444.34 &         5437.85 &     5439.55 \\
 ARITEMETIC EXPRESSION 7  &    794.52 &  10986.54 &                  1514.19 &                  {\textbf{457.27}} &             605.58 &       3415.58 &         3427.72 &     3460.51 \\
 ARITEMETIC EXPRESSION 8  &    868.02 &  10030.57 &                  2090.34 &                  {\textbf{468.90}} &             717.37 &       3646.37 &         3636.16 &     3641.61 \\
 ARITEMETIC EXPRESSION 9  &   4535.89 &  38031.66 &                 13882.09 &                 {\textbf{2823.60}} &            5093.13 &      18729.37 &        18695.13 &    18697.83 \\
 ARITEMETIC EXPRESSION 10 &   1721.44 &  13483.72 &                  4166.67 &                 {\textbf{1005.85}} &            1911.69 &       7004.21 &         7013.96 &     6996.58 \\
 RANDOM EXPRESSION 1      &   2449.01 &  55822.93 &                  3454.41 &                 {\textbf{2112.71}} &            2350.52 &      12849.99 &        12853.38 &    12858.24 \\
 RANDOM EXPRESSION 2      &   1354.32 &   4737.32 &                  1465.94 &                 1441.29 &            1440.93 &      {\textbf{1323.64}} &         1327.30 &     1324.32 \\
 RANDOM EXPRESSION 3      &   3382.98 &  69212.48 &                  5741.97 &                 {\textbf{2945.31}} &            3689.13 &      17524.99 &        17485.55 &    17498.33 \\
 RANDOM EXPRESSION 4      &   3067.34 &  56868.36 &                  4927.13 &                 {\textbf{3063.33}} &            3593.86 &      17996.21 &        17975.26 &    17973.13 \\
 RANDOM EXPRESSION 5      &   5870.61 & 121944.99 &                  9633.84 &                 {\textbf{5146.98}} &            6081.06 &      32546.81 &        32577.18 &    32540.89 \\
 RANDOM EXPRESSION 6      &   8956.69 & 210992.49 &                 17702.72 &                 {\textbf{7106.15}} &            9176.62 &      57433.00 &        57359.91 &    57322.07 \\
 RANDOM EXPRESSION 7      &   4285.56 &  74364.46 &                  6026.55 &                 {\textbf{3914.23}} &            4445.95 &      21011.76 &        21001.10 &    20992.49 \\
 RANDOM EXPRESSION 8      &   1452.89 &  34528.74 &                  1422.12 &                 {\textbf{1353.50}} &            1494.55 &       8719.67 &         8722.33 &     8717.14 \\
 RANDOM EXPRESSION 9      &   {\textbf{4105.70}} &  62012.19 &                  4355.59 &                 4287.66 &            4502.97 &      19235.06 &        19222.82 &    19235.69 \\
 RANDOM EXPRESSION 10     &   4951.93 & 116824.13 &                  9718.61 &                 {\textbf{4481.14}} &            5487.71 &      32010.81 &        32014.66 &    31985.01 \\
\bottomrule
\end{tabular}

\end{adjustbox}
    \caption{Time for each expression (1{,}000 $\times$ 10{,}000 runs) in ms on Linux. The relative timings are similar on different platforms and OS. The complete results can be found on our \href{https://geometryprocessing.github.io/intervals/}{github page}.}
    \label{table:linux}
\end{table*}

\begin{figure}
    \parbox{0.02\linewidth}{~}\hfill\hfill
    \parbox{0.32\linewidth}{\centering Single Operations}\hfill
    \parbox{0.32\linewidth}{\centering Arithmetic Expressions}\hfill
    \parbox{0.32\linewidth}{\centering Random Expressions}\par
    \parbox{0.02\linewidth}{\centering\rotatebox{90}{\sffamily\scriptsize{Time(ms)}}}\hfill\hfill
    \parbox{0.32\linewidth}{\centering\sffamily\scriptsize
    \includegraphics[width=\linewidth]{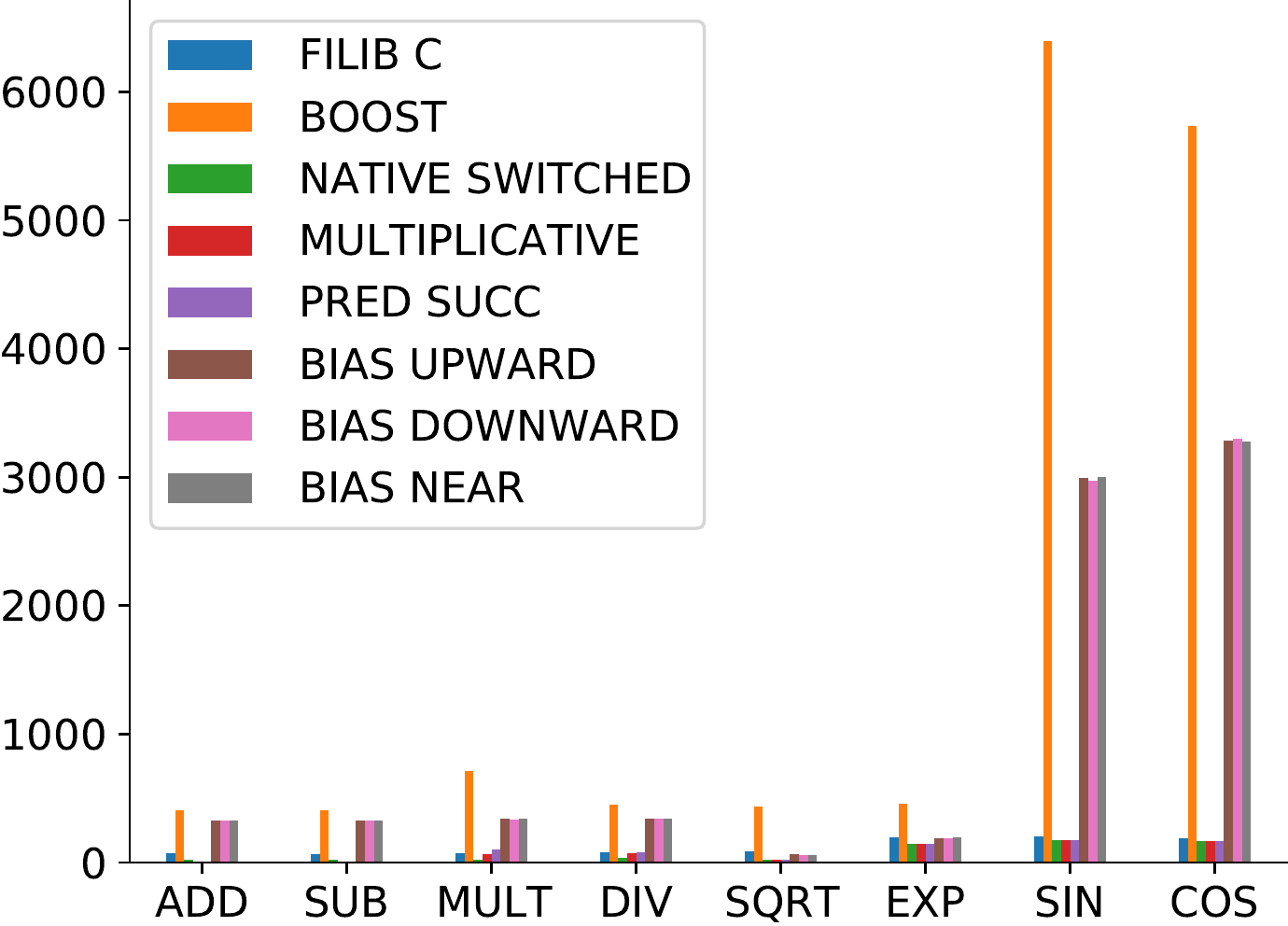}}\hfill
    \parbox{0.32\linewidth}{\centering\sffamily\scriptsize
    \includegraphics[width=\linewidth]{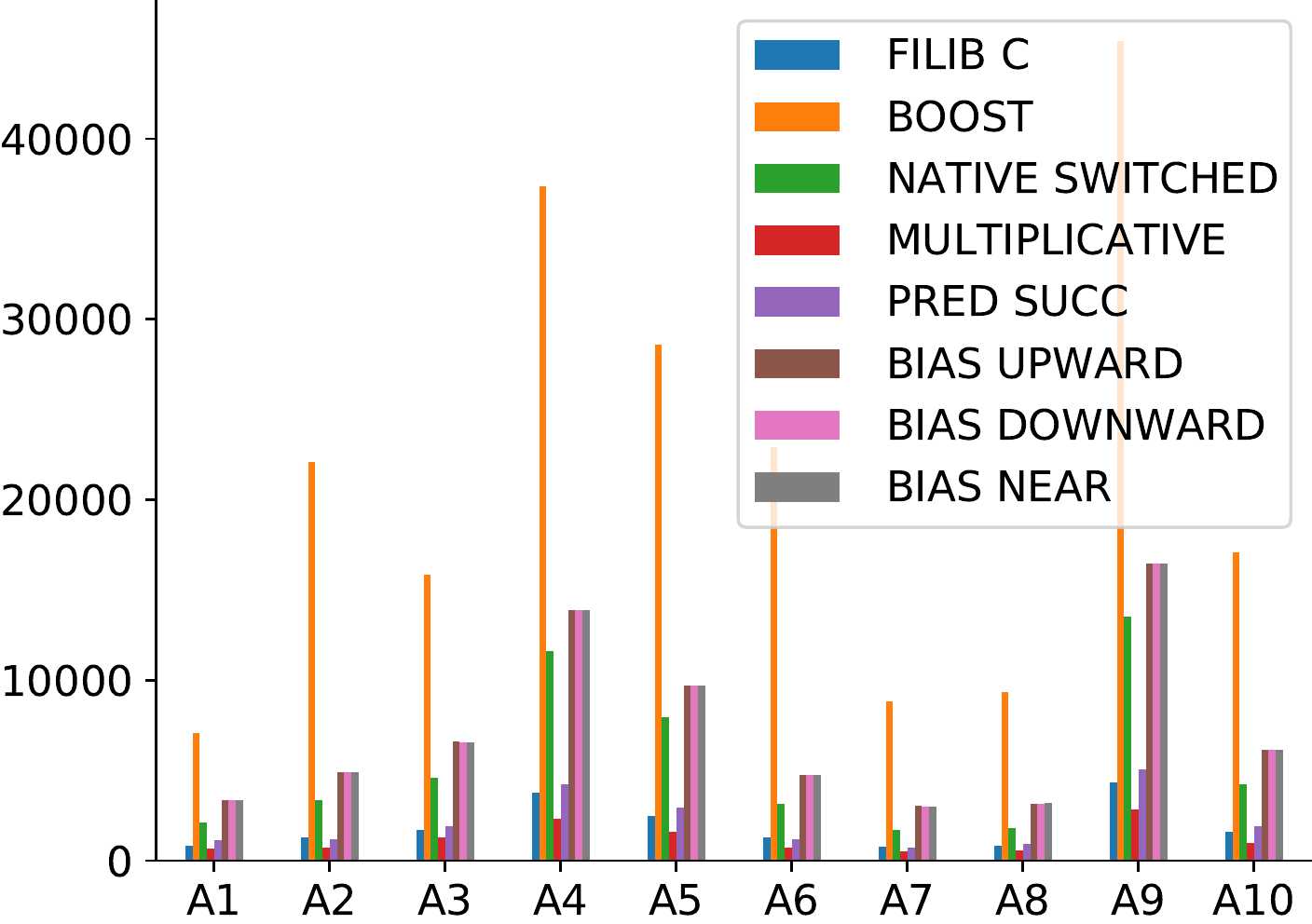}}\hfill
    \parbox{0.32\linewidth}{\centering\sffamily\scriptsize
    \includegraphics[width=\linewidth]{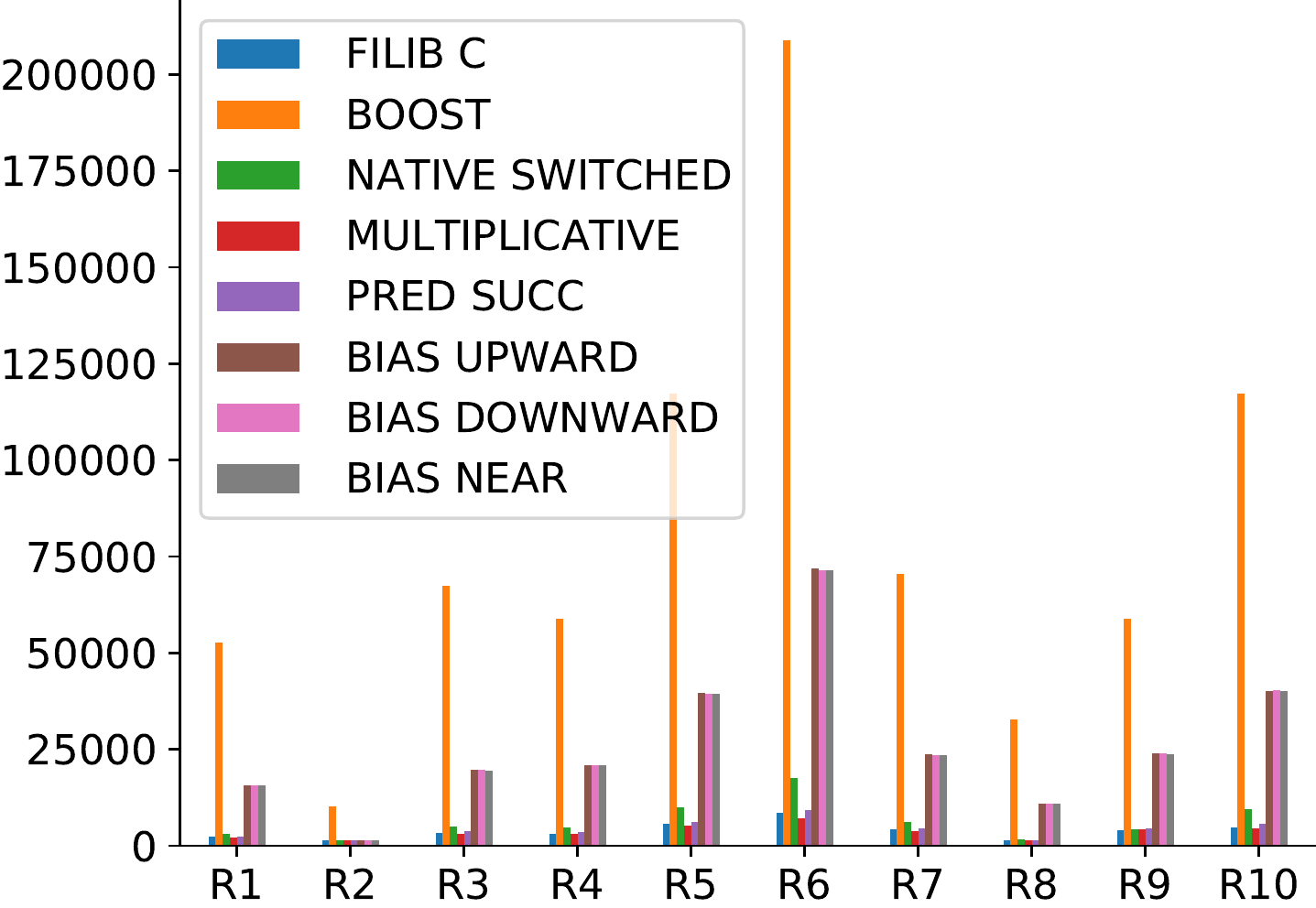}}\par
    \caption{Time for each expression (1{,}000 $\times$ 10{,}000 runs) in ms on Linux.}
    \label{fig:performance_linux}
\end{figure}

We show the accumulated time in milliseconds for each expression on the Linux platform since the relative performance across platforms is similar. We also highlight the fastest method for each expression.
From Table~\ref{table:linux}, we see that \boost has the worst performance on all of the 28 expressions, followed by BIAS, then filib. While filib++'s \nativeswitched mode also sets rounding mode for basic arithmetic operations, it is highly optimized and is significantly faster than the other two libraries.

Within filib++, the performance of the three modes on basic operations is comparable. However, for more complex arithmetic expressions, \nativeswitched mode is consistently the slowest, likely because it changes the rounding mode for each operations. The multiplicative mode is always the fastest, while \predsucc method is between the two. 
Since filib++ ignores interval mode when computing transcendental functions, the performance on $\operatorname{sqrt}$, $\exp$, $\sin$, $\cos$ are similar. As a result, when computing more complicated expressions, multiplicative mode remains the fastest one among three modes and among all the interval types, followed by \predsucc mode, then \nativeswitched mode. Although the speed of filib++ can drop below filib or even BIAS on some expressions, the relative difference is minimal.

\subsection{Consistency and Portability}

While \boost can be deployed on all platforms we test on, it does not produce consistent results due to its system specific rounding modes.

\begin{figure}
    \parbox{0.02\linewidth}{~}\hfill\hfill
    \parbox{0.48\linewidth}{\centering Mac}\hfill
    \parbox{0.48\linewidth}{\centering Linux}\par
    \parbox{0.02\linewidth}{\centering\rotatebox{90}{\sffamily\scriptsize{Percentage}}}\hfill\hfill
    \parbox{0.48\linewidth}{\centering\sffamily\scriptsize
    \includegraphics[width=\linewidth]{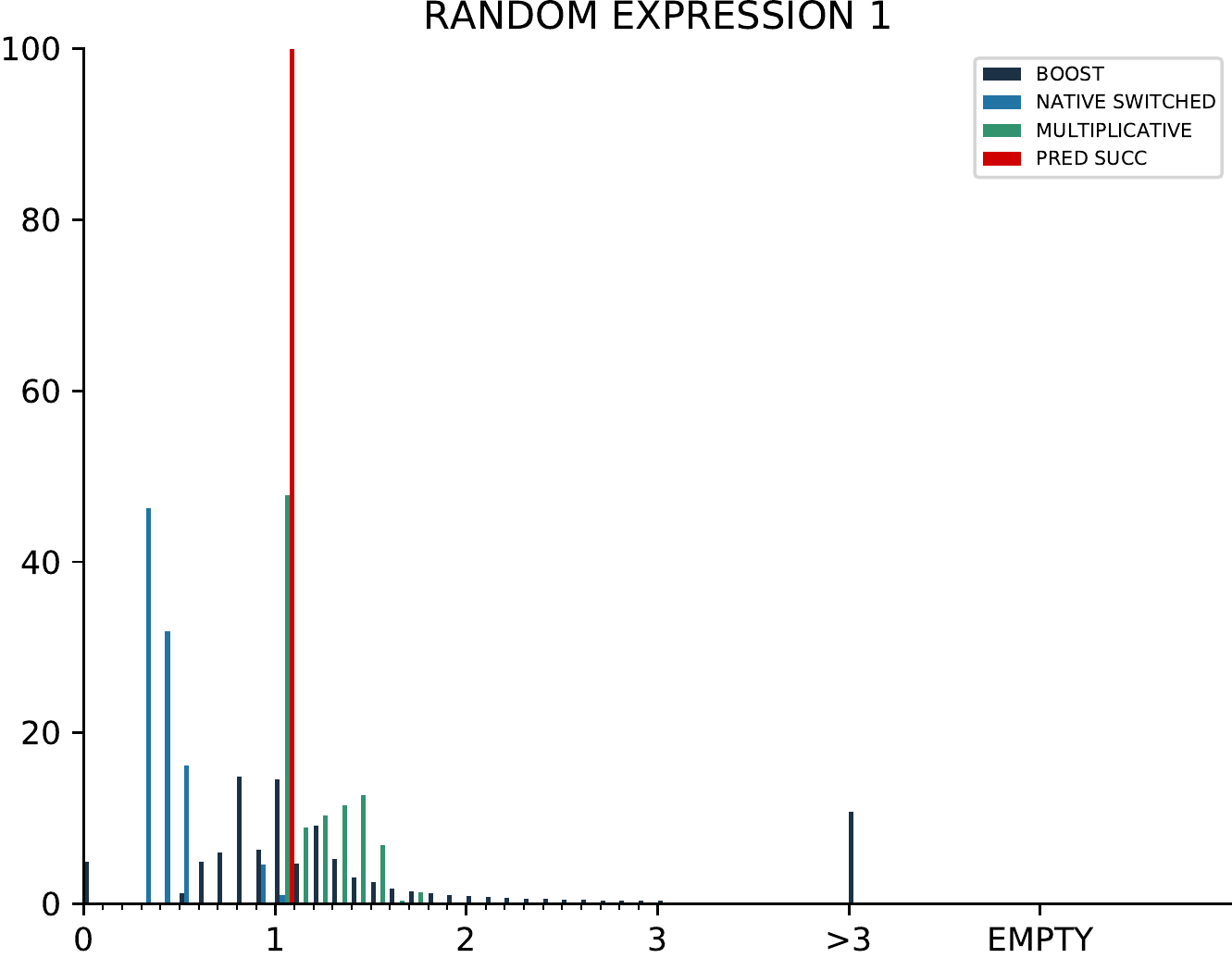}}\hfill
    \parbox{0.48\linewidth}{\centering\sffamily\scriptsize
    \includegraphics[width=\linewidth]{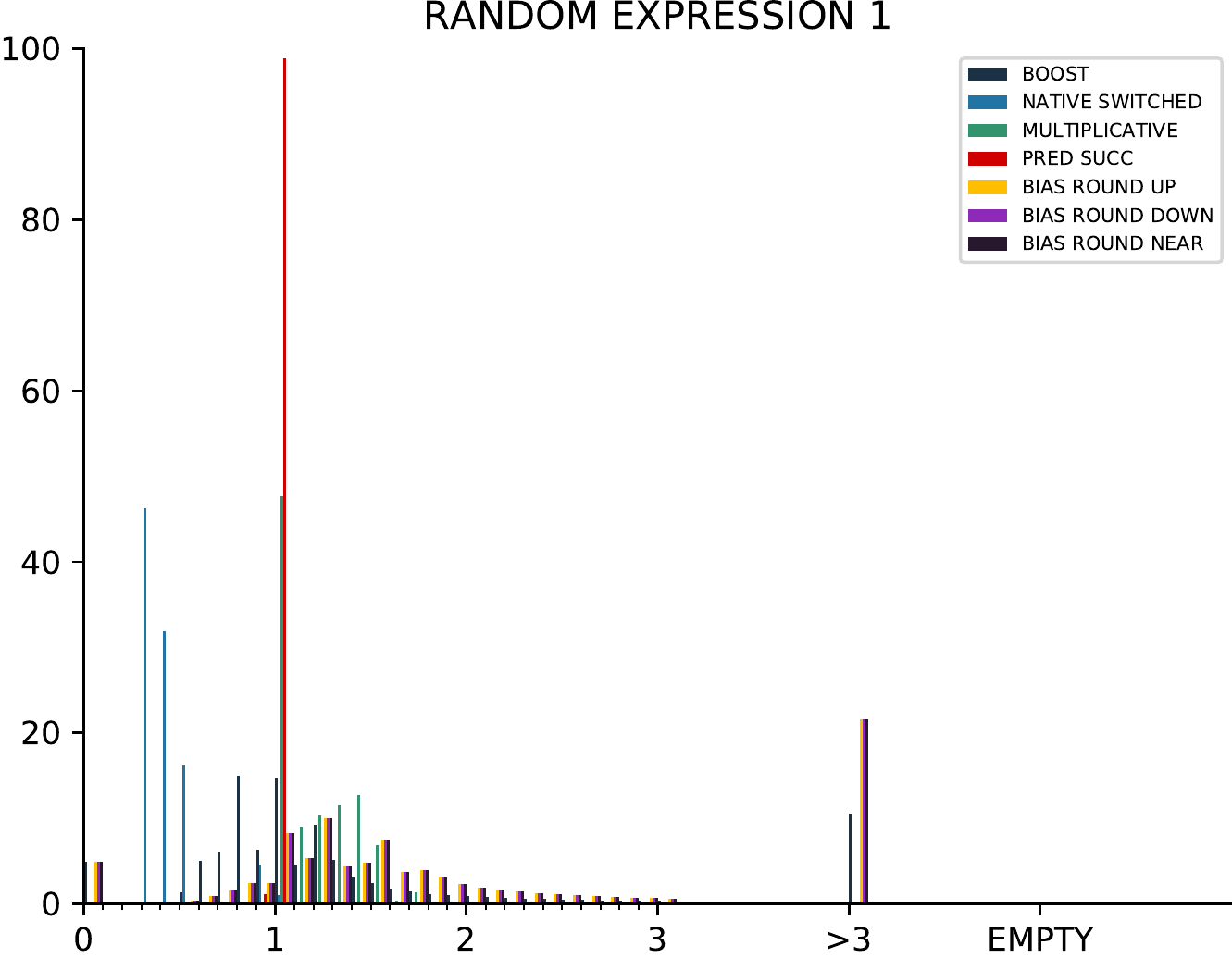}}\par
    \caption{Distribution of each library's interval size on expression \ref{expression:random1}, normalized.}
    \label{fig:random1}
\end{figure}

While filib++ does well in terms of both correctness and speed, it does not produce the same result across different platforms: we found that it produces different results on the Linux platform. As seen in Figure \ref{fig:random1}, the \predsucc's distribution of interval size differs from Linux to Mac.  Additionally, its portability is limited due to the lack of updates since 2011 and the use of autoconf to generate the makefile
\footnote{
\href{http://www2.math.uni-wuppertal.de/wrswt/software/filib.html}{filib++ source}, last updated in 2011.
}

BIAS is not maintained
\footnote{
\href{http://www.ti3.tu-harburg.de/keil/profil/index_e.html}{BIAS source}, last updated in 2009,
}
and currently does not compile out of the box on modern windows and macOS versions. We thus only tested it on Linux.

\subsection{Application on Continuous Collision Detection Queries}

As a further benchmark of correctness, we integrate three interval libraries in the continuous collision detection (CCD) benchmark of \cite{wang2020large}. The CCD benchmark features two interval based algorithms to detect collisions along a continuous linear trajectory. Both the univariate and multivariate interval-based CCD perform interval-based bisection root finding \cite{snyder1992interval} and use interval arithmetic to compute an estimate of the codomain of a function. The correctness of the interval arithmetic ensures that no false negatives (no collision is reported when there is a collision) occur and smaller interval size helps to reduce the number of false positives (a collision is reported when there is no collision).

\begin{figure}\centering
\parbox{.60\linewidth}{\centering
    \parbox{0.02\linewidth}{\centering\rotatebox{90}{\sffamily\scriptsize{Time(us)}}}\hfill\hfill
    \parbox{0.48\linewidth}{\centering\sffamily\scriptsize
    \includegraphics[width=\linewidth]{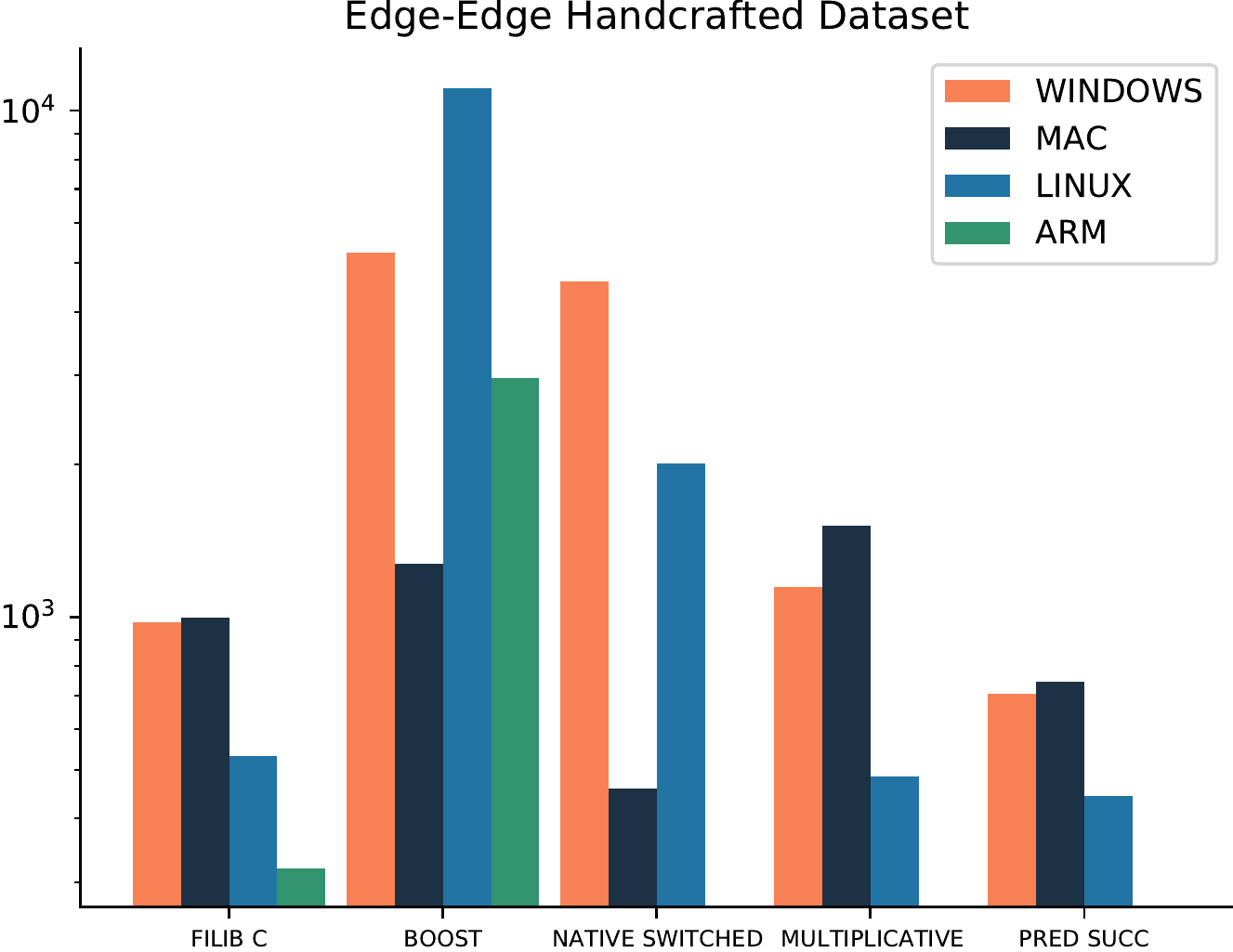}}\hfill
    \parbox{0.48\linewidth}{\centering\sffamily\scriptsize
    \includegraphics[width=\linewidth]{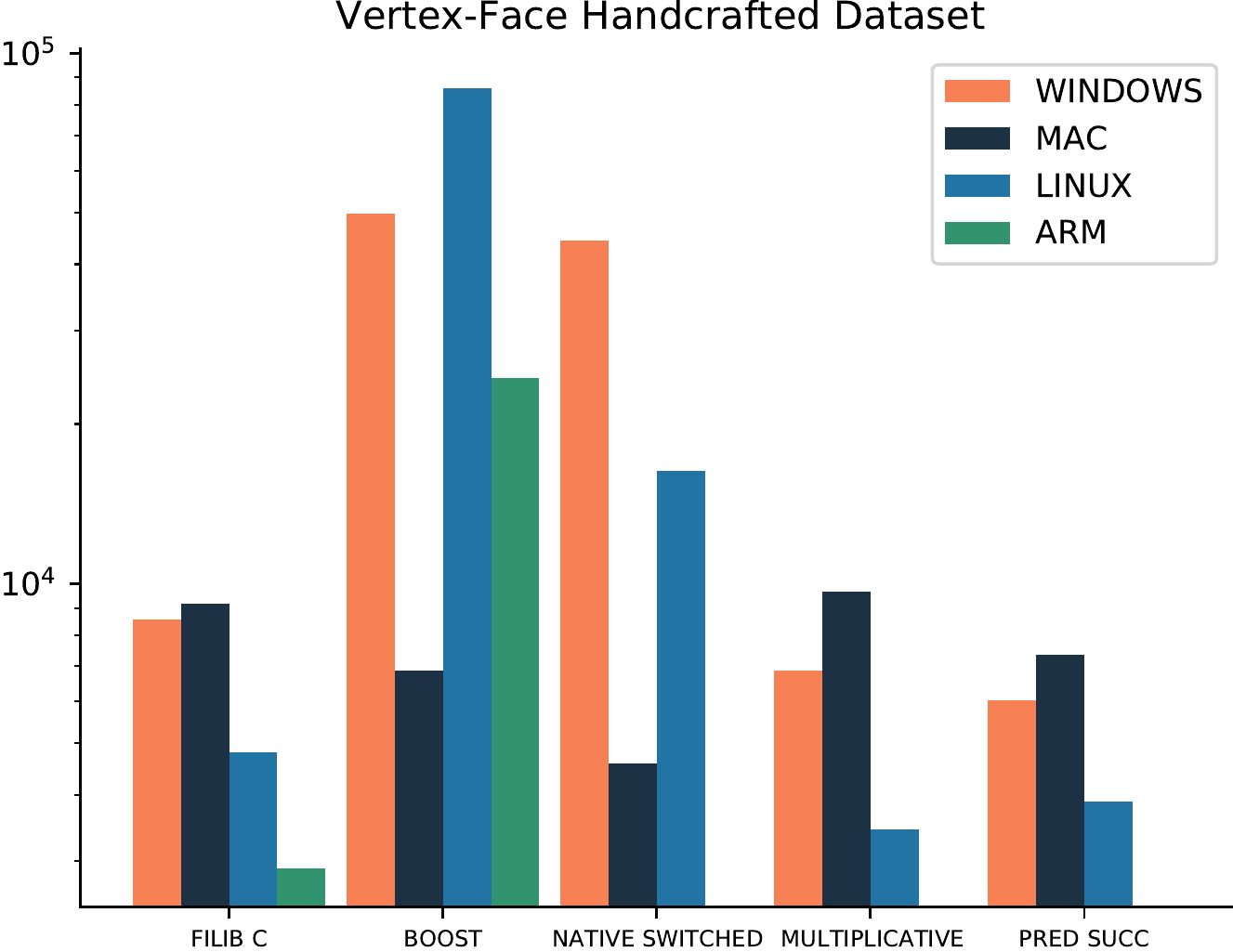}}\par
    \parbox{0.02\linewidth}{\centering\rotatebox{90}{\sffamily\scriptsize{Time(us)}}}\hfill\hfill
    \parbox{0.48\linewidth}{\centering\sffamily\scriptsize
    \includegraphics[width=\linewidth]{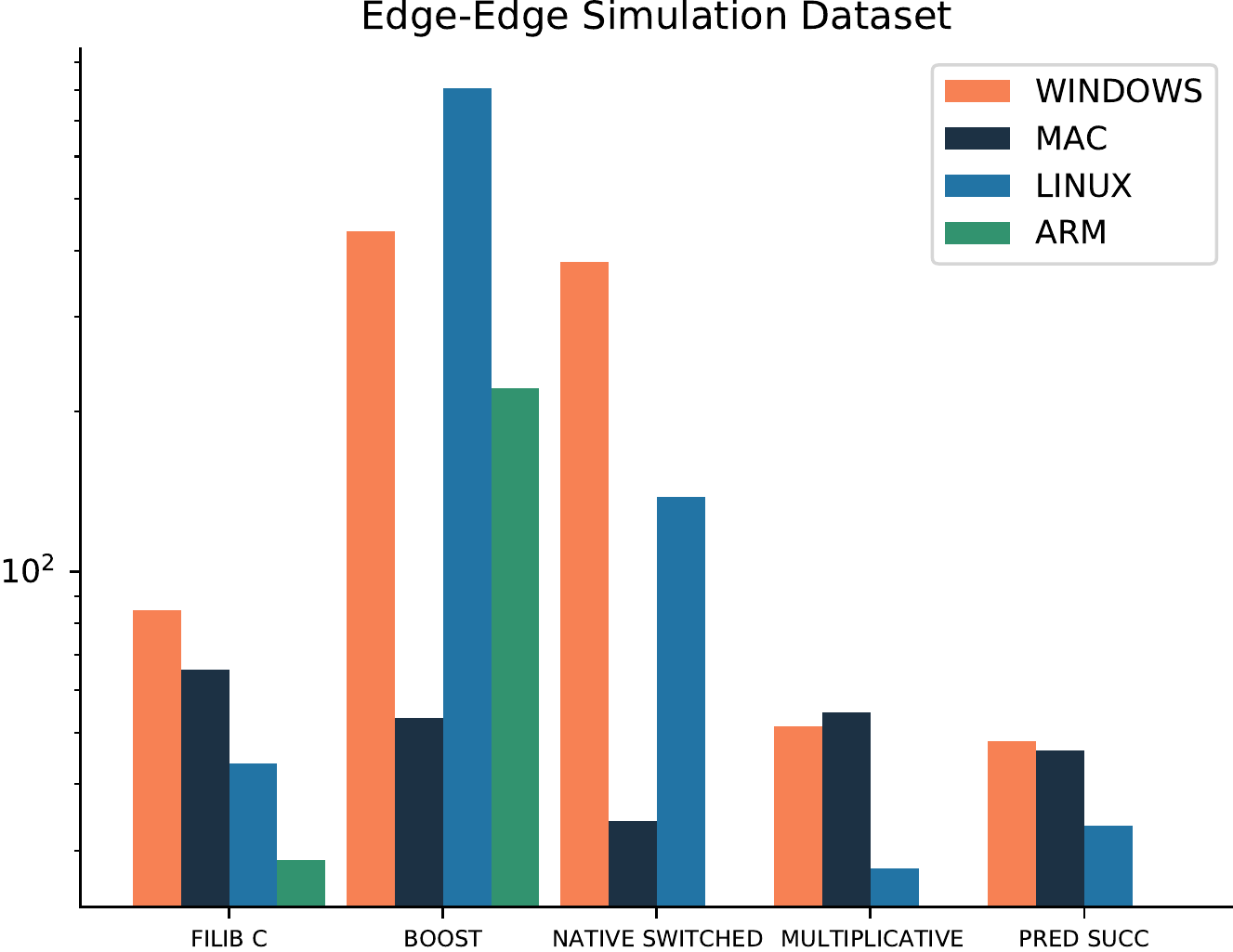}}\hfill
    \parbox{0.48\linewidth}{\centering\sffamily\scriptsize
    \includegraphics[width=\linewidth]{graphs/ccd_time/UnivariateIntervalRootFinder_Edge-Edge_Handcrafted_Datasettime.pdf}}\\[1.5em]
}
    \caption{Average time of each query using different interval types on different platform in univariate interval root finder test.}
    \label{graph:ccd_time_uni}
\end{figure}

\begin{figure}\centering
\parbox{.60\linewidth}{
    \parbox{0.02\linewidth}{\centering\rotatebox{90}{\sffamily\scriptsize{Time(us)}}}\hfill\hfill
    \parbox{0.48\linewidth}{\centering\sffamily\scriptsize
    \includegraphics[width=\linewidth]{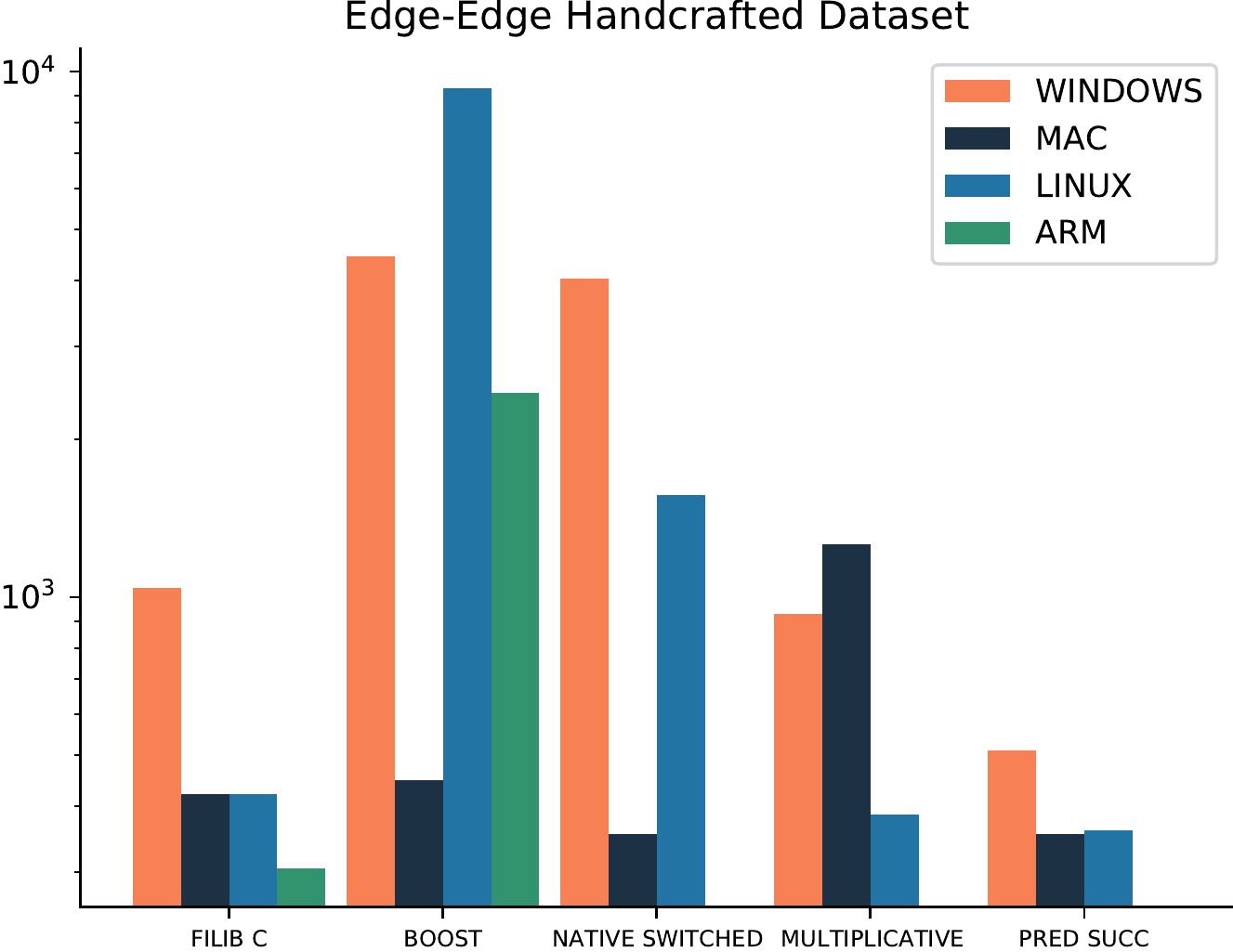}}\hfill
    \parbox{0.48\linewidth}{\centering\sffamily\scriptsize
    \includegraphics[width=\linewidth]{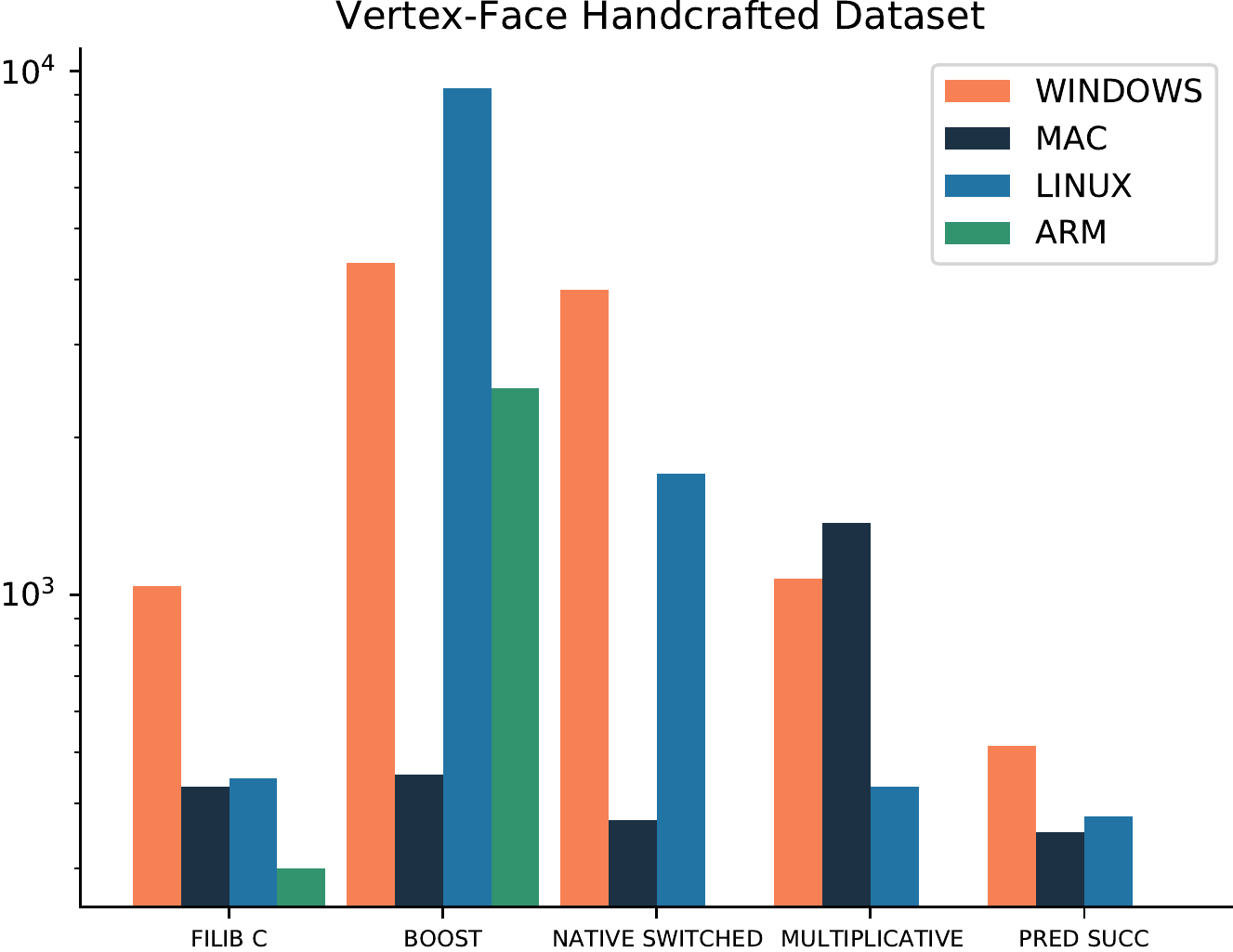}}\par
    \parbox{0.02\linewidth}{\centering\rotatebox{90}{\sffamily\scriptsize{Time(us)}}}\hfill\hfill
    \parbox{0.48\linewidth}{\centering\sffamily\scriptsize
    \includegraphics[width=\linewidth]{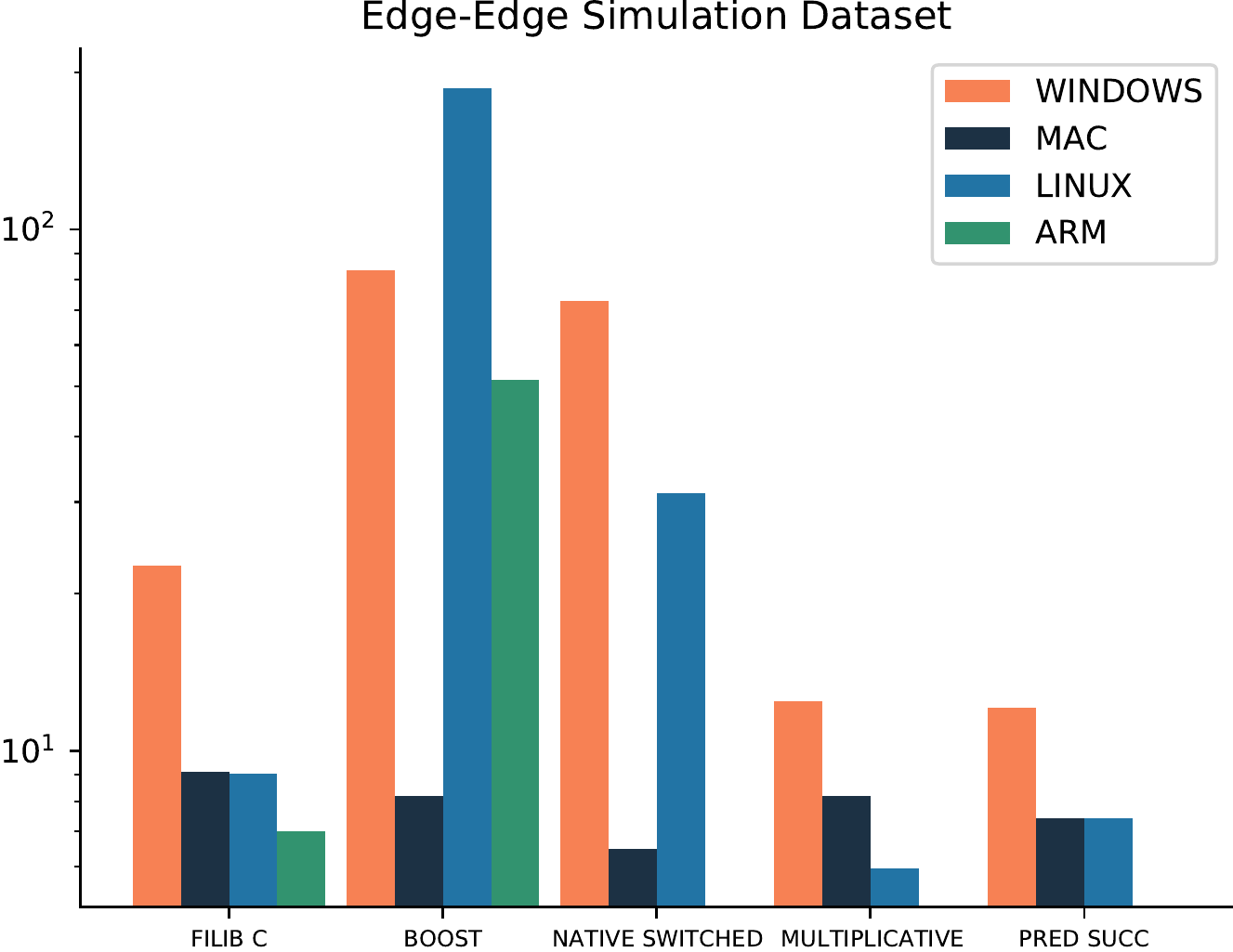}}\hfill
    \parbox{0.48\linewidth}{\centering\sffamily\scriptsize
    \includegraphics[width=\linewidth]{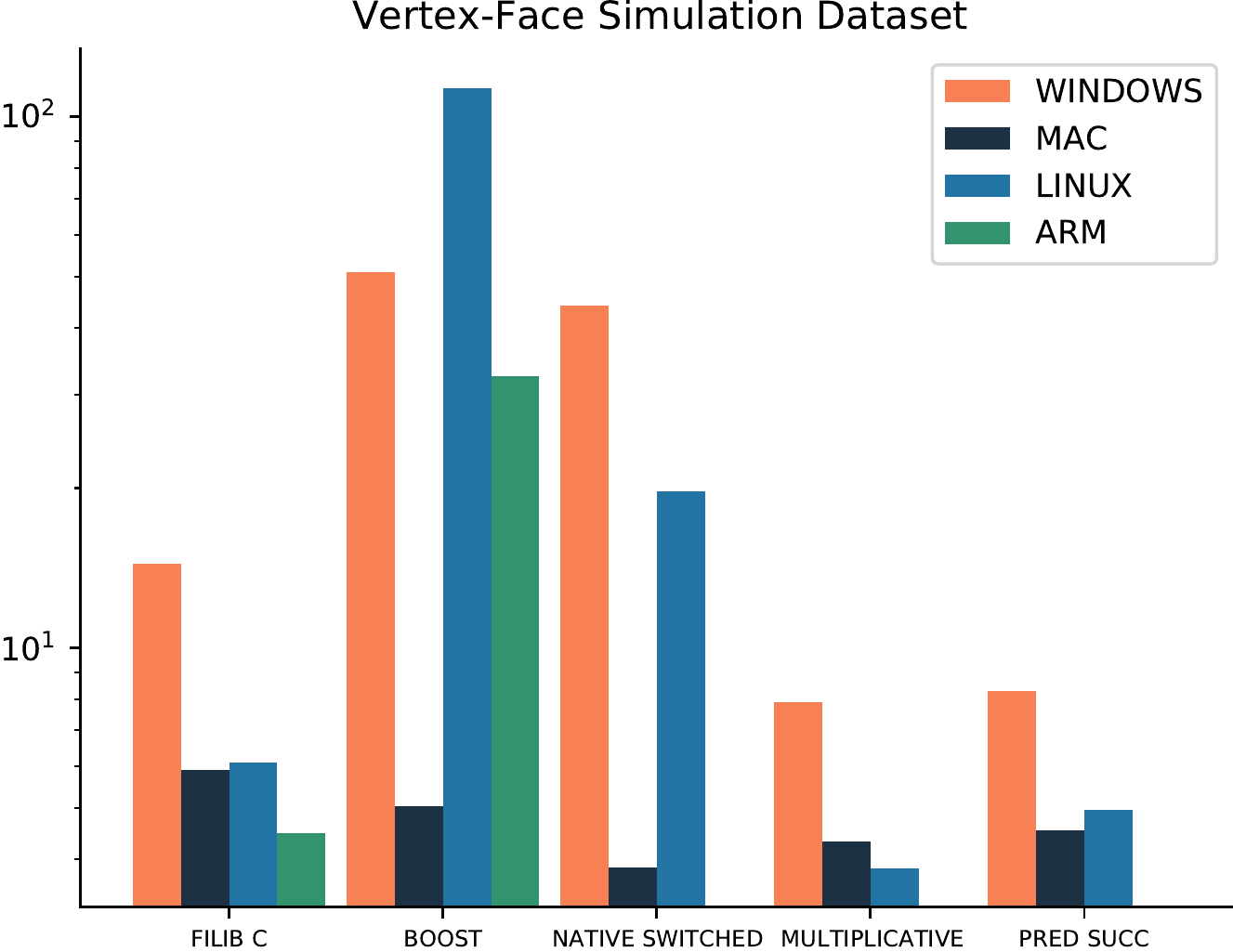}}\par
}
    \caption{Average time of each query using different interval types on different platform in multivariate interval root finder test.}
    \label{graph:ccd_time_multi}
\end{figure}

Timing-wise (Figure \ref{graph:ccd_time_uni}, \ref{graph:ccd_time_multi}), all libraries are in a similar ballpark, with the exception of \boost being slightly slower than the others in certain tests. 
\begin{figure}\centering
\parbox{.60\linewidth}{
    \parbox{0.02\linewidth}{\centering\rotatebox{90}{\sffamily\scriptsize{Count}}}\hfill\hfill
    \parbox{0.48\linewidth}{\centering\sffamily\scriptsize
    \includegraphics[width=\linewidth]{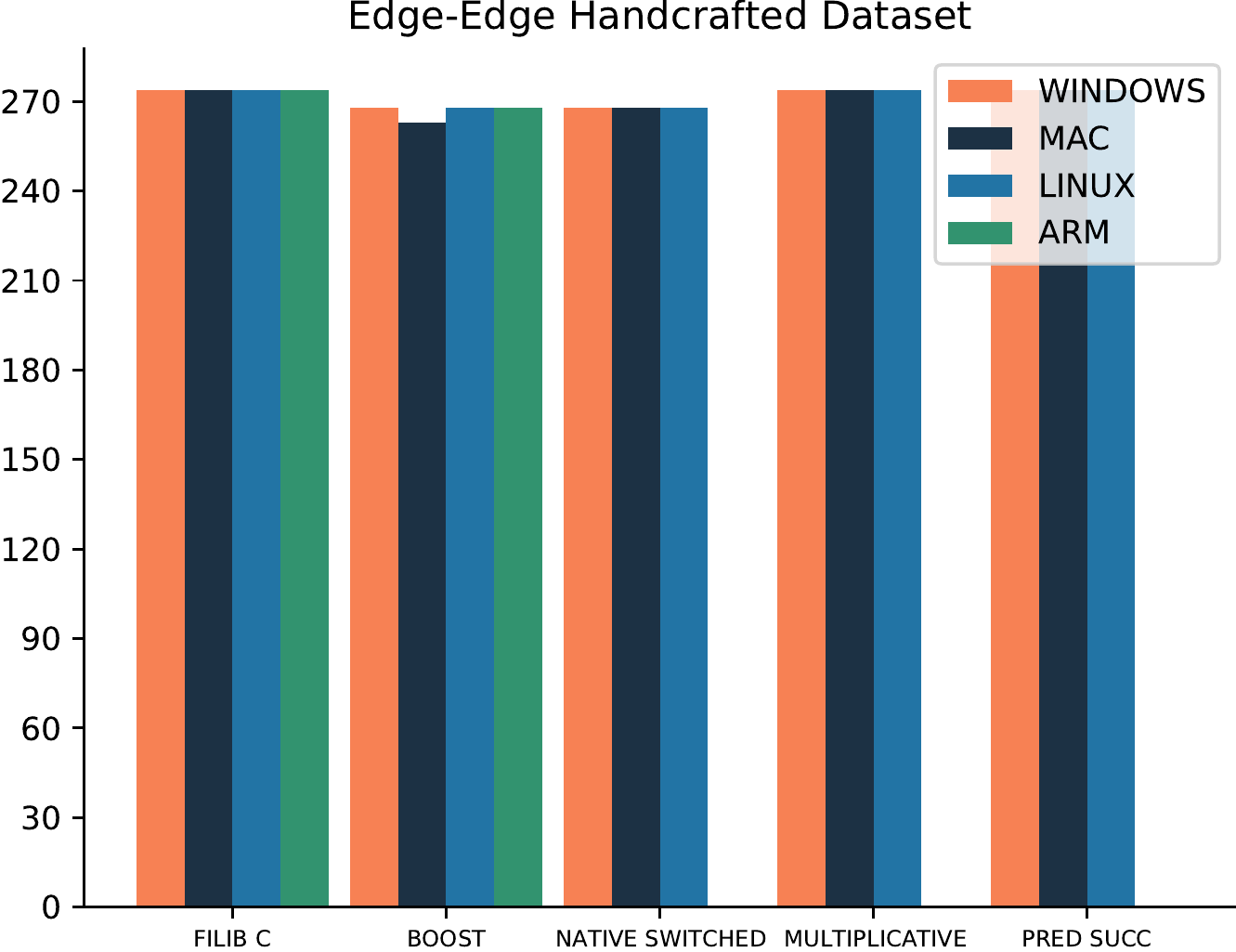}}\hfill
    \parbox{0.48\linewidth}{\centering\sffamily\scriptsize
    \includegraphics[width=\linewidth]{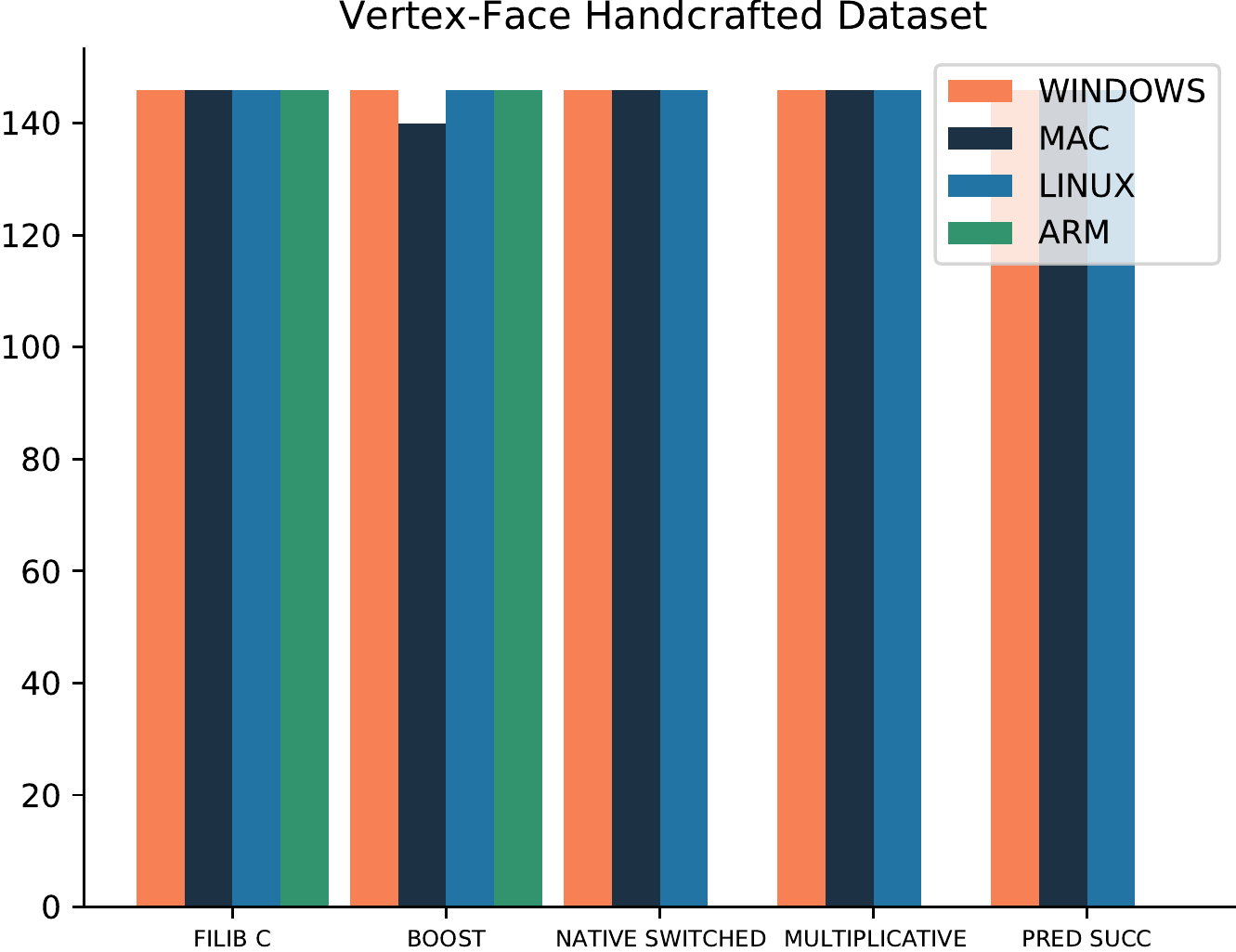}}\par
    \parbox{0.02\linewidth}{\centering\rotatebox{90}{\sffamily\scriptsize{Count}}}\hfill\hfill
    \parbox{0.48\linewidth}{\centering\sffamily\scriptsize
    \includegraphics[width=\linewidth]{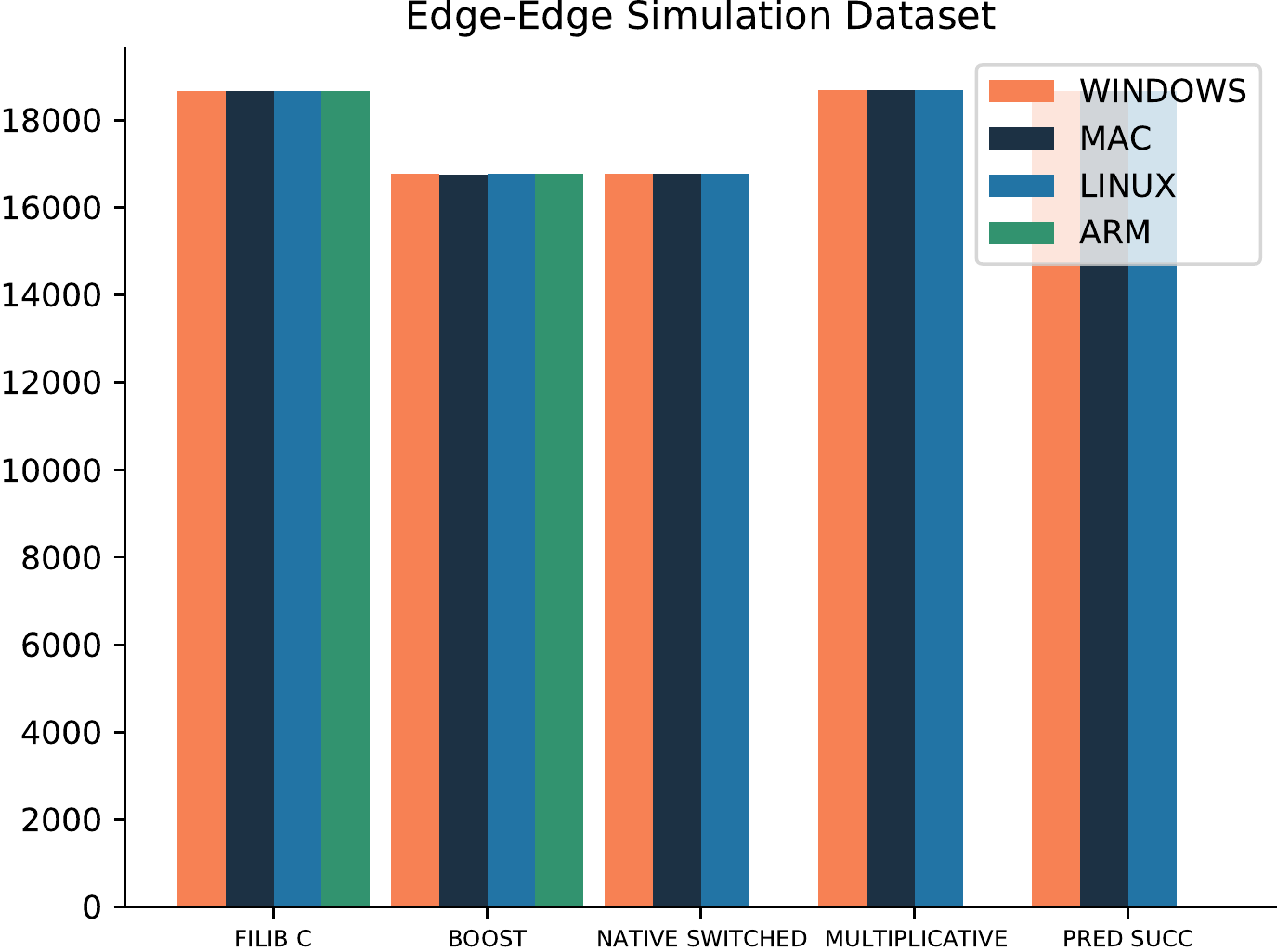}}\hfill
    \parbox{0.48\linewidth}{\centering\sffamily\scriptsize
    \includegraphics[width=\linewidth]{graphs/ccd_false_positive/UnivariateIntervalRootFinder_Edge-Edge_Handcrafted_Datasetfalse_positives.pdf}}\\[1.5em]
}
    \caption{Number of false positives using different interval types on different platform in univariate interval root finder test.}
    \label{graph:ccd_false_positive_uni}
\end{figure}

\begin{figure}\centering
\parbox{.60\linewidth}{\centering
    \parbox{0.02\linewidth}{\centering\rotatebox{90}{\sffamily\scriptsize{Count}}}\hfill\hfill
    \parbox{0.48\linewidth}{\centering\sffamily\scriptsize
    \includegraphics[width=\linewidth]{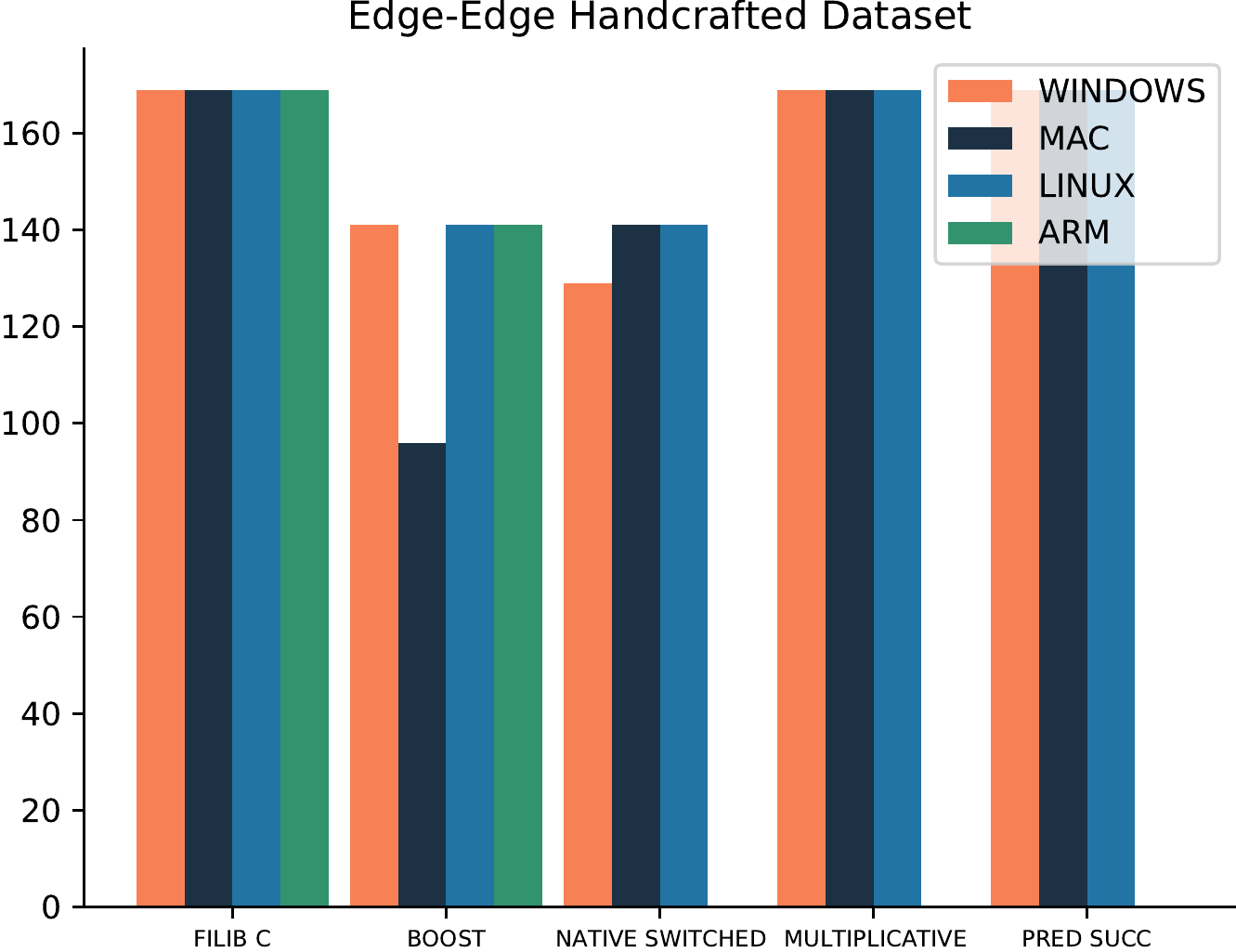}}\hfill
    \parbox{0.48\linewidth}{\centering\sffamily\scriptsize
    \includegraphics[width=\linewidth]{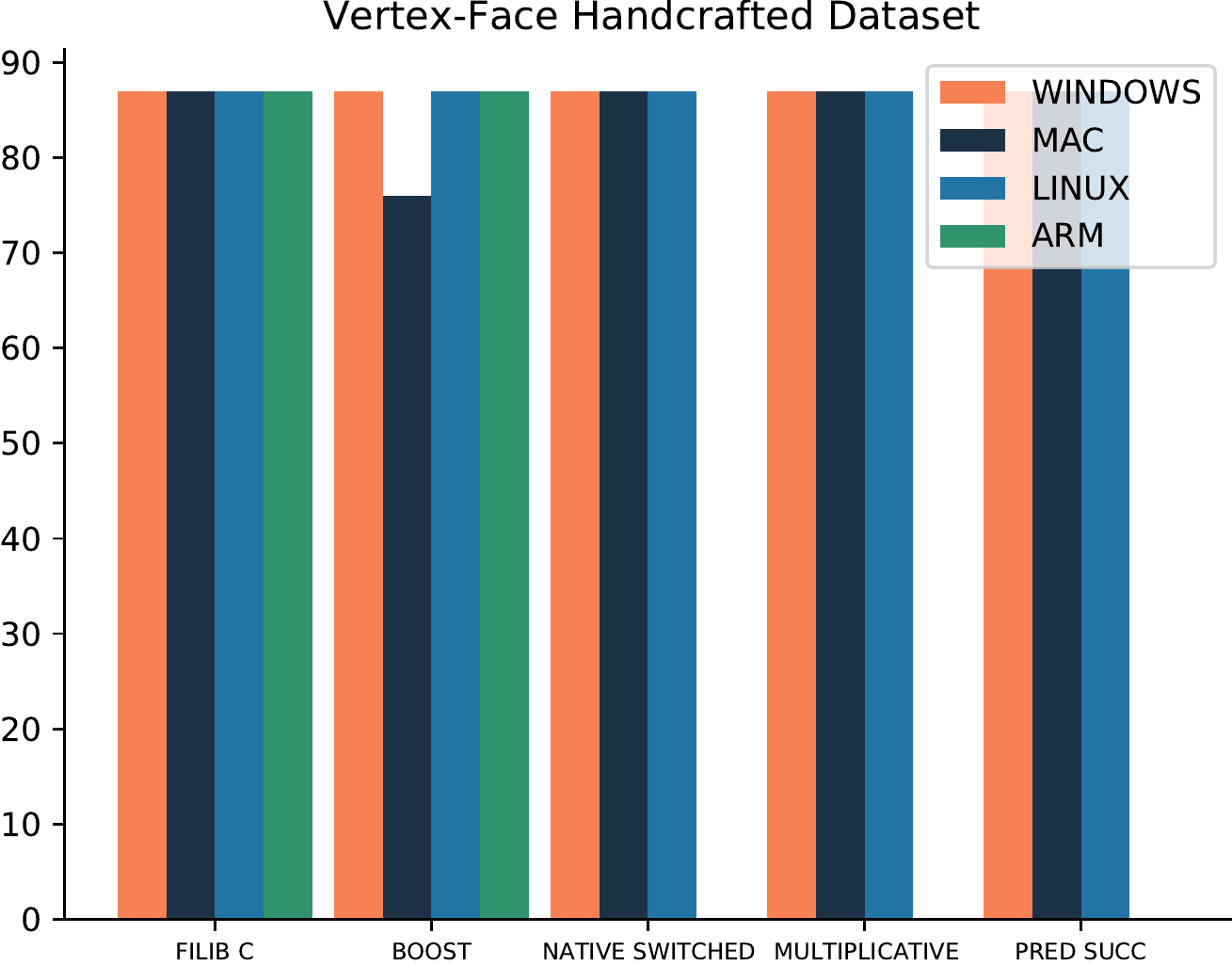}}\par
    \parbox{0.02\linewidth}{\centering\rotatebox{90}{\sffamily\scriptsize{Count}}}\hfill\hfill
    \parbox{0.48\linewidth}{\centering\sffamily\scriptsize
    \includegraphics[width=\linewidth]{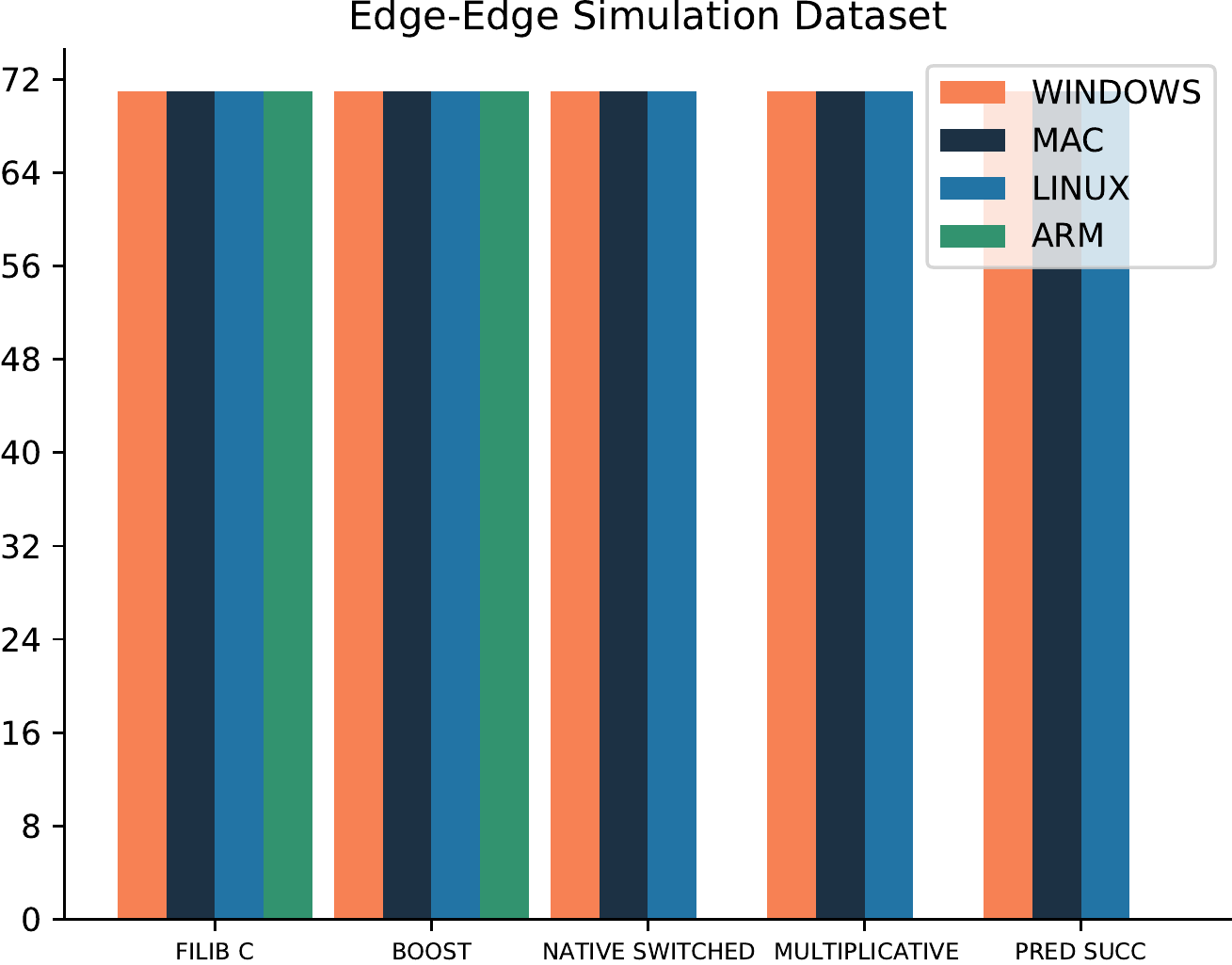}}\hfill
    \parbox{0.48\linewidth}{\centering\sffamily\scriptsize
    \includegraphics[width=\linewidth]{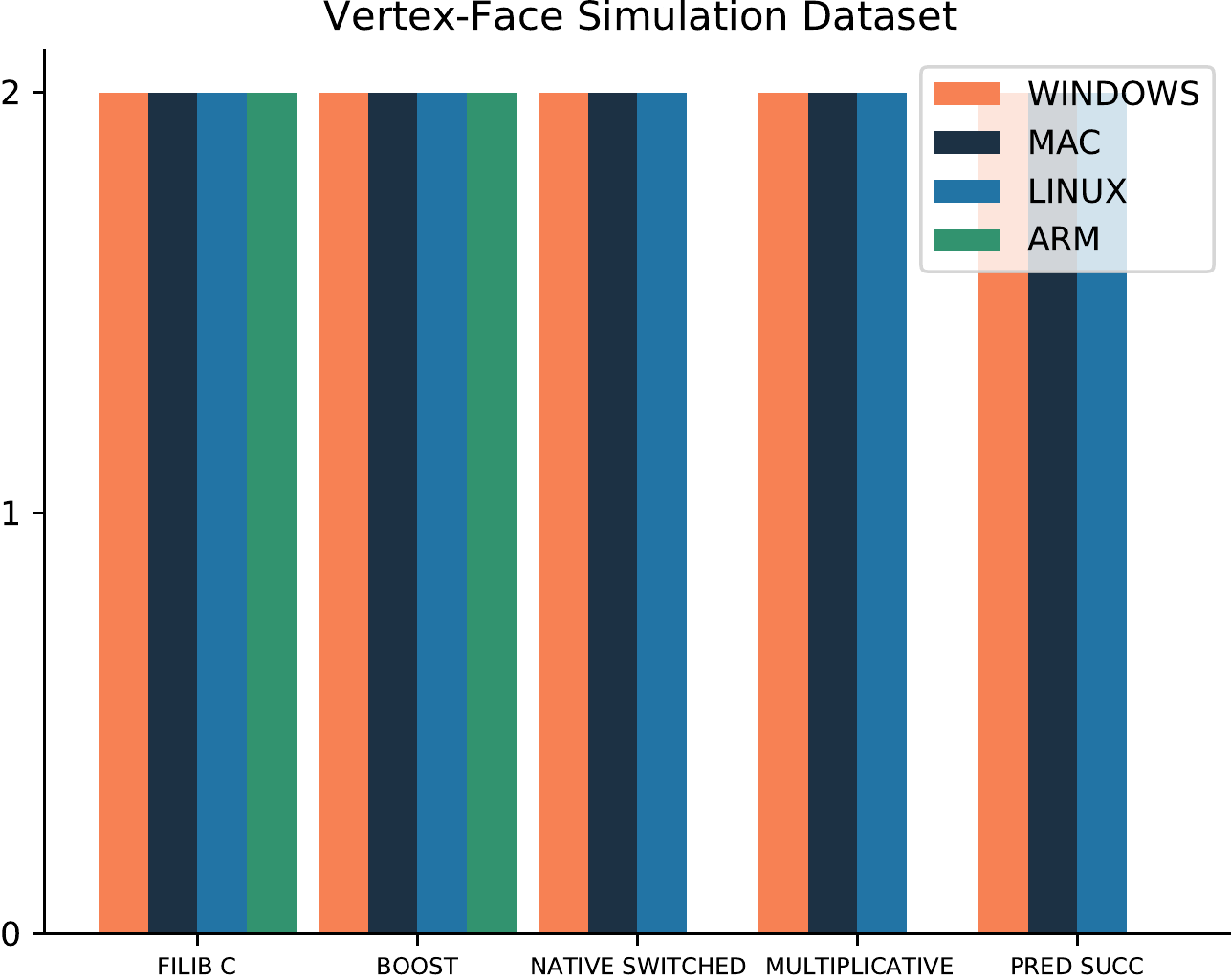}}\par
}
    \caption{Number of false positives using different interval types on different platform in multivariate interval root finder test.}
    \label{graph:ccd_false_positive_multi}
\end{figure}
None of the libraries produces false negatives in this benchmark. The number of false positives varies as expected, as the intervals are different (Figure \ref{graph:ccd_false_positive_uni}, \ref{graph:ccd_false_positive_multi}). It is concerning to see that \boost and filib++'s \nativeswitched produce different numbers of false positives on different architectures. filib, filib++'s \predsucc, and filib++'s \multiplicative method produce consistent results across all operating systems and architectures.

\section{Conclusion}
\label{sec:conclusion}

In this paper, we designed a benchmark that tests interval libraries for correctness, interval size, speed, and consistency. Using our benchmark we evaluated four interval libraries: filib, filib++, \boost, and BIAS (Table \ref{fig:teaser}). We also provide the complete results along with all the expressions on our github page.
\footnote{
\href{https://geometryprocessing.github.io/intervals/}{https://geometryprocessing.github.io/intervals/}
}


In our study, filib is the only library that is correct, consistent, portable, and efficient. We believe it is the best option between the libraries we tested. To make deployment on multiple platforms easier, we provide a copy of the library with a modern cmake build system on github.
\footnote{
\href{https://github.com/txstc55/filib}{https://github.com/txstc55/filib}.
}

\clearpage

\bibliographystyle{unsrt}  
\bibliography{references}  

\appendix
\label{sec:appendix}
\section{Expressions} \label{sec:expressions}
Expressions that contains only arithmetic operations:

\begin{equation}
    \frac{a  (a + b  c)}{(b + c  d)} - \frac{d  \left(e + {f}/{g}\right)}{(g + h)} - \frac{i}{j}
\label{expression:arith2}
\end{equation}

\noindent Expressions that contains transcendental functions:

\begin{equation}
    \cos\left(\left(\cos\left(\cos(f)+\exp\left(\frac{d}{c}\right)\right)\right) \left(\sin\left(\sqrt{e}+a+b-\sqrt{d+c}\right)\right)\right)
    \label{expression:random1}
\end{equation}

\begin{equation}
    \exp\left(\sqrt{\exp\left(\sqrt{\exp\left(\sqrt{a}\right)}\right)}\right)
    \label{expression:random2}
\end{equation}

\begin{equation}
\begin{split}
    \exp\left(\frac{\sqrt{{\exp\left(\cos\left({a}/{d}\right)\right)}/{\exp\left(\cos\left(\sqrt{f}\right)\right)}}}
    {\sqrt{{\cos(\cos(\cos(c)))}/{\sqrt{\sin(\cos(b))}}}}\right)
\end{split}
\label{expression:random9}
\end{equation}


\end{document}